\documentclass[11pt,a4paper,reqno]{amsart}
\usepackage{amsmath}
\usepackage{amsfonts}
\usepackage{amssymb}
\usepackage{amsthm}
\usepackage{newlfont}
\usepackage{a4}
\usepackage{enumerate}
\usepackage[pdftex]{graphicx,color}
\usepackage{backref}

\newsavebox{\Imagebox}
\usepackage{pstricks}

\theoremstyle{definition}

\theoremstyle{remark}

\usepackage[centertags]{amsmath}
\usepackage{graphicx}
\usepackage{amsfonts}
\usepackage{amssymb}
\usepackage{amsthm}
\usepackage{bbm}
\usepackage{newlfont}
\usepackage{a4}
\usepackage{enumerate}

\newlength{\defbaselineskip}
\newcommand{\setlinespacing}[1]%
           {\setlength{\baselineskip}{#1 \defbaselineskip}}

\allowdisplaybreaks

\newcommand{\q}{\quad}
\renewcommand{\epsilon}{\varepsilon}

\newcommand{\la}{\lambda}

\renewcommand{\rho}{\varrho}
\renewcommand{\phi}{\varphi}

\newcommand{\R}{{\mathbb{R}}}

\newcommand{\N}{{\mathbb N}}

\newcommand{\Z}{\mathbb{Z}}

\newcommand{\abs}[1]{\left\vert#1\right\vert}

\allowdisplaybreaks

\begin{document}

\title[]{Four families of orthogonal polynomials of $C_2$  and\\ symmetric and antisymmetric generalizations of \\ sine and cosine functions}


\author{L. Motlochov\'{a}$^{1}$}
\author{J. Patera$^{2}$}

\begin{abstract}\

Four families of generalizations of trigonometric functions, denoted
$\cos^\pm_{(\lambda,\mu)}(x,y)$ and $\sin^\pm_{(\lambda,\mu)}(x,y)$, were recently introduced. In the paper the functions are transformed into four families of orthogonal polynomials depending on two variables. Recurrence relations for construction of the polynomials are presented. Orthogonality relations of the four families of polynomials are found together with the appropriate weight fuctions. Tables of the lowest degree polynomials are shown. Numerous trigonometric-like identities are found.

Two of the four families of functions are identified as the functions encountered in the Weyl character formula for the finite dimensional irreducible representations of the compact Lie group $Sp(4)$. The other two families of functions seem to play no role in Lie theory so far in spite of their analogous `good' properties.
\end{abstract}
\date{\today}

\maketitle
\noindent
$^1$ Departement de math\'ematiques et de statistique, Universit\'e de Montr\'eal, C.~P.~6128, succ. Centre-ville, Montr\'eal, H3C\,3J7, Qu\'ebec, Canada; motlochova@dms.umontreal.ca
\\
$^2$ Centre de recherches math\'ematiques, Universit\'e de Montr\'eal, C.~P.~6128, succ. Centre-ville,
Montr\'eal, H3C\,3J7, Qu\'ebec, Canada; patera@crm.umontreal.ca\\

\section{Introduction}

Four families of special functions and four related families of orthogonal polynomials in two real variables are studied in the paper. They are the symmetric and antisymmetric generalizations of cosine and sine functions denoted here as $\cos^\pm$ and $\sin^\pm$. The simplicity of the functions \eqref{trigfunctions} combined with their continuous and discrete orthogonality should make them particularly amenable to application. 

Orthogonal polynomials were studied intensively in recent years using various approaches frequently inspired by orthogonal polynomials in one variable. Many such polynomials are described in the literature \cite{Koorn,suetin,Mac2,Mac3}.  In the absence of additional constraints, truly higher dimensional generalizations are hidden in the vast number of possibilities of defining orthogonal polynomials of more than one variable \cite{suetin,Dunkl,kow1,kow2}. In $2D$ \cite{Koorn}, such constraints were provided by requiring that polynomials of two variables be simultaneous eigenvectors of two differential operators.

The idea of our construction of the polynomials in its simplest realization is rater transparent: Consider the characters of the finite dimensional irreducible representations of a compact simple Lie group of rank $n$. Declare the characters $\chi_{\omega_1},\dots,\chi_{\omega_n}$ of the fundamental representations to be the polynomial variables. Any other character is then a polynomial in the lowest $n$ of them. It is constructed recursively by a judicious choice of products of the lower characters with the $n$ basic ones and their decomposition into the sum of irreducible characters. Orthogonality and other properties of such polynomials follow from the corresponding properties of characters.

The idea of our construction remains valid when the characters are replaced by their constituents. That is the Weyl group invariant orbit functions ($C$-functions here), or the Weyl group skew-invariant orbit functions ($S$-functions here) that appear in the numerator of the Weyl character formula, or the `hybrid' functions $C^-$ and $S^+$ defined in this paper.

Our starting point are the special functions of $n$ real variables \cite{KPtrig}, which we specialize to two variables, and from those functions we construct the 2-variable polynomials. In order to do that, we substitute the variables \eqref{variables} into the functions of \eqref{functions}. The functions thus become polynomials in $X$ and $Y$. The recurrence relations of subsections {\ref{c+recurrence}-- {\ref{s-recurrence} allow one to find the polynomials to any desired degree. Several lowest degree polynomials for each of the four families are shown in Appendix~2.

The immediate goal of our paper is to describe properties of such functions and their polynomials in details impossible in the $n$-variable generality. In this way we obtain only a limited number of the polynomials of 2 variables found in the literature, but our approach has other advantages besides the possibility of generalization to any number of variables obvious from \cite{KPtrig}. Let us now point out the main advantages.

The most important conclusion of the paper is the link between the special functions, generalizations of the symmetric and antisymmetric sine and cosine, and their polynomials to the compact simple Lie group of type $C_2$ (also called $O(5)$ or $Sp(4)$). Relating the polynomials to the compact simple Lie groups, rather than to the root systems of such groups \cite{Mac3}, provides a discretization of the polynomials through the discretization of the special functions \cite{MP06}. Discrete orthogonality of the functions carries over to discrete orthogonality of the polynomials on lattices of any density and of symmetry dictated by the underlying Lie group. See examples in \cite{MP2010,PZ05,PZ06}. Moreover a generalization to the simple Lie group of any rank $>2$ is straightforward and leads to polynomials of any number of variables \cite{MP2010}.

The most curious result of the paper is the existence of two new families of orthogonal special functions, called here $C^-$ and $S^+$, besides the families of $C$ and $S$ types which are known from the Weyl character formula \cite{BS}. Analogous new families of special functions are to be studied elsewhere for every compact simple Lie group with two root lengths, the number of variables being the rank of the Lie group. The new families of functions do not exist for the simple Lie groups with roots of equal length, i.e. types $A_n$, $D_n$, $E_6$, $E_7$, and $E_8$.

Previous results \cite{HP,HMP} about the $C_2$-functions are extended in this paper in several ways:
\smallskip
\newline
- One-to-one correspondence is shown between functions of the families $\cos^+$, $\sin^-$ and the $C$- and $S$-functions appearing in the Weyl formula for characters of finite dimensional irreducible representations of the compact simple Lie algebra $C_2$ of the Lie groups $Sp(4)$ and $O(5)$. The correspondence is a result of a straightforward linear substitution of variables \eqref{substitution}. The fundamental weight basis and its dual in the case of $C$- and $S$-functions are transformed into the orthogonal basis of $\cos^+$ and $\sin^-$ functions.
\smallskip
\newline
- The same substitution of variables reveals that the functions $\cos^-$ and $\sin^+$ correspond to two families of functions, denoted here as $C^-$ and $S^+$, whose role in Lie theory has apparently not been noted before. The functions of the new families have all the `good' properties of the $C$- and $S$-functions, which are called here the $C^+$- and $S^-$-functions. The notable distinction between the four families of functions is their behaviour at the boundary $\partial F$ of the isosceles right triangle $F$, the domain of their orthogonality in the real Euclidean space $\R^2$. 
\smallskip
\newline
- A method of constructing polynomials of two variables from the special functions is developed and applied to finding some lowest degree polynomials for each family. It can be understood as a non-linear substitution of variables linking $F\subset\R^2$ and its image $\mathfrak F\subset\R^2$ after the substitution of variables \eqref{functions}. Both $F$ and $\mathfrak F$ are shown in Fig.~\ref{integrationdomains}.
\smallskip
\newline
- Continuous orthogonality of the polynomials, when integrated over $\mathfrak F$, is demonstrated. An appropriate weight fuction is determined for each family\footnote{The discrete orthogonality of the polynomials, when summed up over a finite set of points, is described elsewhere for each family and for several low rank simple Lie groups.}. It is a consequence of the orthogonality of the irreducible characters of any compact simple Lie group when they are integrated over the maximal torus, and of the symmetries of our functions under the action of the Weyl group. 
\smallskip
\newline
- Numerous trigonometric-like identities are found involving generalized trigonometric functions of the four families. Many of them are shown in  Appendix~1.

\medskip

The history of the generalization of trigonometric functions considered here can be traced in the three recent papers \cite{KPtrig,HP,HMP}. The $C$-functions of the simple Lie algebra $C_2$ were considered in \cite{PZ05,KP06}, and the $S$-functions of $C_2$ are described in \cite{PZ06,KP07}. The polynomial representation of the functions \eqref{trigfunctions} has not been considered elsewhere. In general, the study of the orthogonal polynomials of $C$-and $S$-functions originated in \cite{NPT}. Their exploitation in polynomial approximations \cite{mhaskar} is another possibility. Simultaneous zero values of polynomials of certain degrees have to be identified and the corresponding cubature formulas used \cite{MP2010,Xu, Dunkl}. Computational aspects of such applications remain unexplored besides the few examples in  \cite{HP,HMP}.

In Section~2, the functions $\cos^\pm_{(\lambda,\mu)}(x,y)$ and $\sin^\pm_{(\lambda,\mu)}(x,y)$ are introduced and their congruence class is defined. The structure of the decomposition of their products is shown in Table~\ref{products}. Changing from the orthogonal basis to bases inherent to the $C_2$ symmetry in Section~3 shows that the functions \eqref{trigfunctions} are in fact, up to a normalization, the $W$-invariant functions of the simple Lie group $C_2$. Two of the functions are well known in Lie theory, the other two are new. The polynomials are described in Section~4. The functions \eqref{functions} to be represented as polynomials in the same variables are considered first. Curiously, the structure of three of the four families of functions resembles the Weyl character formula. In fact using the variables of the previous Section, it is easy to verify that the function $\sin^-$ is the character function of $C_2$. The variables $X$ and $Y$ are then introduced, the lowest few polynomials of $\cos^+_{(\lambda,\mu)}(x,y)$ are constructed, and the full set of recurrence relations is written for each family. The functions studied here are shown in Section~5 to be eigenfunctions of the Laplace operator and of another algebraically independent 4-th degree operator. In orthogonal coordinates this fact is visible from \eqref{trigfunctions}. In the polynomial variables both operators are considerably more complicated. Section~6 examines the continuous orthogonality of the polynomials, known for the functions \eqref{trigfunctions}, and rewritten here in the basis $\{X,Y\}$. Unexpectedly, the kernel--Jacobian of the orthogonality integrals known from Lie theory factorizes if it `knows' of the existence of the two new families of functions. Section~7 contains the generic identities verified by the functions \eqref{trigfunctions}. Special identities are shown in Appendix~1. The last Section contains some concluding remarks. Appendix~2 consists of eight tables of explicit low degree polynomials, two tables per family. A comparison of the lowest polynomials between different families of \eqref{functions} can easily be made there.

\section{Notations, definitions, and some properties}
We first recall the definitions of functions symmetric and antisymmetric in variables $x$ and $y$, introducing simplified notations,
\begin{equation}\label{trigfunctions}
\begin{alignedat}{2}
c^+_{(\lambda,\mu)}&:=\cos^+_{(\lambda,\mu)}(x,y)=
    \cos(\pi\lambda x)\cos(\pi\mu y)+\cos(\pi\mu x)\cos(\pi\lambda y)
        \qquad&&\text{(\it symmetric cosine)}\\
c^-_{(\lambda,\mu)}&:=\cos^-_{(\lambda,\mu)}(x,y)=
    \cos(\pi\lambda x)\cos(\pi\mu y)-\cos(\pi\mu x) \cos(\pi\lambda y)
         \quad&&\text{(\it antisymmetric cosine)}\\
s^+_{(\lambda,\mu)}&:=\sin^+_{(\lambda,\mu)}(x,y)=
    \sin(\pi\lambda x)\sin(\pi\mu y)+\sin(\pi\mu x)\sin(\pi\lambda y)
         \qquad&&\text{(\it symmetric sine)} \\
s^-_{(\lambda,\mu)}&:=\sin^-_{(\lambda,\mu)}(x,y)=
    \sin(\pi\lambda x)\sin(\pi\mu y)-\sin(\pi\mu x)\sin(\pi\lambda y)
         \qquad&&\text{(\it antisymmetric sine)} 
\end{alignedat}
\end{equation}
Here $x,y\in\R$ and $\lambda,\mu\in\Z$, such that
\begin{equation}
\begin{alignedat}{2}\label{inequalities}
 &\lambda\geq\mu\geq0\qquad\text{for}\quad c^+_{(\lambda,\mu)}\,,
 &\qquad\qquad
 &\lambda > \mu\geq0\qquad\text{for}\quad c^-_{(\lambda,\mu)}\\
 &\lambda\geq\mu > 0\qquad\text{for}\quad s^+_{(\lambda,\mu)}\,,
 &\qquad\qquad
 &\lambda > \mu > 0\qquad\text{for}\quad s^-_{(\lambda,\mu)}
\end{alignedat}
\end{equation}
Note that $c^-$ and $s^-$ are $2\times 2$ determinants, while $c^+$ and $s^+$ are $2\times 2$ permanents \cite{minc}. Assuming that $\lambda$ and $\mu$ are integers, it is advantageous to split the functions within each family into two congruence classes according to the value of the congruence number $\#$ defined as
\begin{equation}\label{congruence}
\#(c^\pm_{(\lambda,\mu)}(x,y))=\#(s^\pm_{(\lambda,\mu)}(x,y))=\lambda-\mu\mod2\,.
\end{equation}
During multiplication of the functions their congruence numbers add up.

Numerous symmetry properties of the polynomials $c^\pm_{(\lambda,\mu)}$ and $s^\pm_{(\lambda,\mu)}$ are shown in \cite{HP} and \cite{HMP} respectively. The functions \eqref{trigfunctions} are clearly defined for all real values of $x$ and $y$ and have continuous derivatives of all degrees. Due to their symmetry properties, we are interested in their values only in the fundamental region $F\subset\R^2$, which is the right isosceles triangle on 
Fig.~ \ref{integrationdomains} with vertices
\begin{equation}\label{triangle}
F:=\{(0,0),\,(1,0),\,(1,1)\}\,.
\end{equation}

Of importance is the distinct behaviour of functions \eqref{trigfunctions} at the boundary $\partial F$ of $F$. Indeed, normal derivatives of $c^+$ functions vanish at $\partial F$, functions $s^-$ are equal to zero at $\partial F$, while $c^-$ and $s^+$ behave differently at long and short sides of the triangle $F$.

Products of pairs of the functions \eqref{trigfunctions} decompose into their sums. 10 different pairs of functions can be multiplied. The decompositions have a rigid structure, which affects which functions may appear. To demonstrate, we reproduce Table~1 of \cite{HMP} as Table~1 here. Actual decompositions of specific products are found in Table~2 of \cite{HMP}.

\begin{table}[h]
\footnotesize
\addtolength{\tabcolsep}{-3pt}

\begin{center}
\begin{tabular}{|c||c|c|c|c|c|c|c|c|c|c|}
\hline
product &$\sin^+\sin^+$  
        &$\sin^+\sin^-$   
        &$\sin^-\sin^-$
        &$\sin^+\cos^+$
        &$\sin^+\cos^-$
        &$\sin^-\cos^+$
        &$\sin^-\cos^-$
        &$\cos^+\cos^+$
        &$\cos^+\cos^-$ 
        &$\cos^-\cos^-$  \\
\hline 
terms   &$\cos^+$ 
        &$\cos^-$
        &$\cos^+$
        &$\sin^+$
        &$\sin^-$
        &$\sin^-$
        &$\sin^+$
        &$\cos^+$ 
        &$\cos^-$ 
        &$\cos^+$   \\
\hline

\end{tabular}
\bigskip
\caption{Structure of the decomposition of the 10 types of products. The second row shows which functions appear in all  the terms of a decomposition.} \label{products}
\end{center}
\end{table}

\section{Orbit functions of $C_2$ and $\cos^\pm$ and $\sin^\pm$}\label{or}
The $C$- and $S$-functions defined by the summation of exponential functions over the Weyl group $W(C_2)$ were recently studied as special functions with many practically useful properties. In Lie theory they are known from the Weyl formula for the characters of irreducible representations of the compact Lie group $C_2$. We show that $C$-functions coincide (up to a scaling factor) with $c^+$-functions defined here, and that $S$-functions similarly coincide with $s^-$-functions. Moreover, we find analogs of $c^-$ and $s^+$ that were not noted in Lie theory before. We describe their properties in parallel with the more familiar $C$- and $S$-functions. See \cite{neworb} for a study of general simple Lie groups not only of type $C_2$.
  
An orbit of $W(C_2)$ consists of all the distinct points (weights) obtained from the dominant point $\underline{v}$ by repeated application of reflections $r_{\alpha_1}$ and $r_{\alpha_2}$ in mirrors orthogonal to the simple roots $\alpha_1$, $\alpha_2$ of $C_2$, according to the formula
$$
r_{\alpha_j}\underline{v}=\underline{v}-
\frac{2\langle\alpha_j,\underline{v}\rangle}{\langle\alpha_j,\alpha_j\rangle}\alpha_j\,.
$$
Assume that the $\alpha_1$ is the shorter of the two simple roots of $C_2$, i.e.
$$
\langle\alpha_1,\alpha_1\rangle=1\,,\qquad
\langle\alpha_2,\alpha_2\rangle=2\,,\qquad
\langle\alpha_1,\alpha_2\rangle=-1\,.
$$ 

In this paper we set $\underline v=(v_1,v_2)=v_1\omega_1+v_2\omega_2$, $(v_1,v_2\in\Z^{\geq0})$, and 
 $\underline\theta=(\theta_1,\theta_2)=\theta_1\check\omega_1+\theta_2 \check\omega_2$, $(\theta_1, \theta_2\in\R)$.
Recall that the scalar product of two vectors, one given in the $\omega$-basis and the other in the dual $\check\omega$-basis of $C_2$, is calculated as follows \cite{KP06}.
$$
\langle\underline v,\underline\theta\rangle
   =\langle(v_1\omega_1+v_2\omega_2),(\theta_1\check\omega_1+\theta_2\check\omega_2)\rangle
   =\sum_{i,j=1}^2v_i\theta_j\langle\omega_i,\check\omega_j\rangle
   =(v_1+v_2)\theta_1+(\tfrac12v_1+v_2)\theta_2\,.
$$

A generic $W(C_2)$-orbit consists of 8 points/weights:
$$
\pm(v_1,v_2)\,,\quad\pm(-v_1,v_1+v_2)\,,\quad\pm(v_1+2v_2,-v_2)\,,\quad
\pm(v_1+2v_2,-v_1-v_2)\,;\qquad v_1,v_2\in\N\,.
$$

The generic $C$- and $S$-functions are denoted in the sequel by $C^+$ and $S^-$ respectively. Together with the new functions $C^-$ and $S^+$, they are defined and identified with the symmetric and antisymmetric trigonometric functions of 2 variables as follows.
\begin{alignat}{2}
C^+_{(\underline v)}(\underline\theta)
    &=\sum_{w\in W(C_2)}e^{2\pi i\langle w\underline v,\underline\theta\rangle}
    &&=4c^+_{(\lambda,\mu)}(x,y)\,,\label{C^+}\\
S^-_{(\underline v)}(\underline\theta)
    &=\sum_{w\in W(C_2)}(-1)^{l_1(w)+l_2(w)}\,e^{2\pi i\langle w\underline v,\underline\theta\rangle}
    &&=-4s^-_{(\lambda,\mu)}(x,y)\,,\label{S^-}\\
C^-_{(\underline v)}(\underline\theta)
    &=\sum_{w\in W(C_2)}(-1)^{l_1(w)}e^{2\pi i\langle w\underline v,\underline\theta\rangle}
    &&=4c^-_{(\lambda,\mu)}(x,y)\,,\label{C^-}\\
S^+_{(\underline v)}(\underline\theta)
    &=\sum_{w\in W(C_2)}(-1)^{l_2(w)}\,e^{2\pi i\langle w\underline v,\underline\theta\rangle}
    &&=-4s^+_{(\lambda,\mu)}(x,y)\,.\label{S^+}
\end{alignat}
Here $l_k(w)$ is the minimal number of times the reflection $r_{\alpha_k}$ is required to generate $w$.

The equalities of the functions in \eqref{C^+} -- \eqref{S^+} are demonstrated using the following change of $\omega$- and $\check\omega$-basis to the orthogonal basis $\{e_1,e_2\}$, where $\langle e_1,e_1\rangle=\langle e_2,e_2\rangle=\frac12$, and the corresponding variables $v_1,\ v_2$ are substituted for $\lambda,\ \mu$, and  $\theta_1,\ \theta_2$ for $x,\ y$,
\begin{equation}\label{substitution}
\begin{alignedat}{4}
\omega_1&=e_1\,,     &\qquad\check\omega_1&=2e_1\,,   &\qquad v_1    &=\lambda-\mu\,,
                     &\qquad\theta_1&=\tfrac12(x-y)\,,\\
\omega_2&=e_1+e_2\,, &\qquad\check\omega_2&=e_1+e_2\,,&\qquad v_2&=\mu\,,
                     &\qquad\theta_2&=y\,,
\end{alignedat}
\end{equation}
In particular, $\#\underline v=\#(v_1,v_2)= v_1\mod2$.

Using our new symbols $C^+$ for $C$-functions and $S^-$ for $S$-functions, we have explicitly, \begin{equation}\label{C+}
\begin{aligned}
C^+_{(v_1,v_2)}(\theta_1,\theta_2)&:=C_{(v_1,v_2)}(\theta_1,\theta_2)
                  =2\cos(2\pi\langle(v_1,v_2), \underline\theta\rangle)
                  +2\cos(2\pi\langle(-v_1,v_1+v_2),\underline\theta\rangle)\\
                 &+2\cos(2\pi\langle(v_1+2v_2,-v_2), \underline\theta\rangle)
                  +2\cos(2\pi\langle(v_1+2v_2,-v_1-v_2), \underline\theta\rangle)\\
                 &=2\left(\cos(\pi((v_1+v_2)2\theta_1+(v_1+2v_2)\theta_2))
                  +\cos(\pi(v_22\theta_1+(v_1+2v_2)\theta_2)) \right.\\
                 &\left.+\cos(\pi((v_1+v_2)2\theta_1+v_1\theta_2))
                  +\cos(\pi(v_22\theta_1+(-v_1)\theta_2)) \right)\,.
 \end{aligned}
  \end{equation}
  
 \begin{equation}\label{S-}
\begin{aligned}
S^-_{(v_1,v_2)}(\theta_1,\theta_2)&:= S_{(v_1,v_2)}(\theta_1,\theta_2)
            =2\cos(2\pi\langle(v_1,v_2), \underline\theta\rangle)
             -2\cos(2\pi\langle(-v_1,v_1+v_2),\underline\theta\rangle)\\
            &-2\cos(2\pi\langle(v_1+2v_2,-v_2), \underline\theta\rangle)
             +2\cos(2\pi\langle(v_1+2v_2,-v_1-v_2), \underline\theta\rangle)\\
            &=2\left(\cos(\pi((v_1+v_2)2\theta_1+(v_1+2v_2)\theta_2))
             -\cos(\pi(v_22\theta_1+(v_1+2v_2)\theta_2)) \right.\\
            &\left.-\cos(\pi((v_1+v_2)2\theta_1+v_1\theta_2))
             +\cos(\pi(v_22\theta_1+(-v_1)\theta_2)) \right)\,.
 \end{aligned}
  \end{equation}
Clearly the $C^+$ and $S^-$ differ by the sign of some terms. More precisely, the sign is negative each time an odd number of reflections $r_{\alpha_1},\ r_{\alpha_2}$ is applied to $\underline v=(v_1,v_2)$. A practically important difference between functions of the two families is their behaviour at the boundary $\partial F$ of $F$. The function $C^+$ is symmetric with respect to reflections across the three sides of $F$, while $S^-$ is antisymmetric. Both are continuous differentiable, therefore normal derivatives of $C^+$ at $\partial F$ are zero, and $S^-$ has zero value at $\partial F$.

The fundamental role of the functions \eqref{C+} and \eqref{S-} in representation theory is asserted in that they comprise the Weyl character formula for the Lie group $C_2$.

The substitutions \eqref{substitution} used in the functions $C^-$ and $S^+$ result respectively in $\cos^-$ and $\sin^+$, which do not play any noted role in Lie theory, as far as we know.
\begin{equation}\label{C-}
\begin{aligned}
C^-_{(v_1,v_2)}(\theta_1,\theta_2)&:=C_{(v_1,v_2)}(\theta_1,\theta_2)
           =2\cos(2\pi\langle(v_1,v_2), \underline\theta\rangle)
           -2\cos(2\pi\langle(-v_1,v_1+v_2),\underline\theta\rangle)\\
          &+2\cos(2\pi\langle(v_1+2v_2,-v_2), \underline\theta\rangle)
           -2\cos(2\pi\langle(v_1+2v_2,-v_1-v_2), \underline\theta\rangle)\\
          &=2\left(\cos(\pi((v_1+v_2)2\theta_1+(v_1+2v_2)\theta_2))
           -\cos(\pi(v_22\theta_1+(v_1+2v_2)\theta_2)) \right.\\
          &\left.+\cos(\pi((v_1+v_2)2\theta_1+v_1\theta_2))
           -\cos(\pi(v_22\theta_1+(-v_1)\theta_2)) \right)\,.
 \end{aligned}
  \end{equation}
Negative signs of terms in \eqref{C-} occur whenever an odd number of reflections $r_{\alpha_1}$ is applied to $\underline v=(v_1,v_2)$.
  
  \begin{equation}\label{S+}
\begin{aligned}
S^+_{(v_1,v_2)}(\theta_1,\theta_2)&:= S_{(v_1,v_2)}(\theta_1,\theta_2)
            =2\cos(2\pi\langle(v_1,v_2), \underline\theta\rangle)
             +2\cos(2\pi\langle(-v_1,v_1+v_2),\underline\theta\rangle)\\
            &-2\cos(2\pi\langle(v_1+2v_2,-v_2), \underline\theta\rangle)
           -2\cos(2\pi\langle(v_1+2v_2,-v_1-v_2), \underline\theta\rangle)\\
            &=2\left(\cos(\pi((v_1+v_2)2\theta_1+(v_1+2v_2)\theta_2))
             +\cos(\pi(v_22\theta_1+(v_1+2v_2)\theta_2)) \right.\\
            &-\left(\cos(\pi((v_1+v_2)2\theta_1+v_1\theta_2))
             -\cos(\pi(v_22\theta_1+(-v_1)\theta_2)) \right)\,.
 \end{aligned}
  \end{equation}
Negative signs of terms in \eqref{S+} occur whenever an odd number of reflections $r_{\alpha_2}$ is applied to $\underline v=(v_1,v_2)$.

\section{Construction of polynomials of $\cos^\pm$ and $\sin^\pm$}

A polynomial is irreducible if it belongs to one of the four families \eqref{trigfunctions} and if it carries specific values of the integers $\la$ and $\mu$, subject to the inequalities \eqref{inequalities} applicable to its family.

Our aim in this section is to represent the following functions,
\begin{equation}\label{functions}
\cos^+_{(\lambda,\mu)}(x,y)\,,\qquad
\frac{\cos^-_{(\lambda,\mu)}(x,y)}{\cos^-_{(1,0)}(x,y)}\,,\qquad
\frac{\sin^+_{(\lambda,\mu)}(x,y)}{\sin^+_{(1,1)}(x,y)}\,,\qquad
\frac{\sin^-_{(\lambda,\mu)}(x,y)}{\sin^-_{(2,1)}(x,y)}\,,
             \qquad\qquad\lambda,\ \mu\in\Z^{\geq0}\,,
\end{equation}
 as polynomials in the variables $X$ and $Y$ of \eqref{variables}. In addition, one must require that $\lambda$ and $\mu$ satisfy the inequalities prescribed in \eqref{inequalities} for each type of function. Furthermore, if we multiply functions $c^+$ and $s^-$ by certain constant $g_{\la\mu}$  given by 
\begin{equation}\label{normalization}
g_{\la\mu}=\begin{cases}\frac12&\la=\mu=0\\2&\la=\mu>0\\2&\la>\mu=0\\4&\text{otherwise}\end{cases}
\end{equation}
we obtain orbit functions defined by the summation of exponential functions over the orbit of the Weyl group instead of the summation over the Weyl group \cite{KP06,KP07}. In \cite{NPT} , there are constructed polynomials of such orbit functions, such polynomials are monomials. Thus, we study here the functions \eqref{functions} multiplied by $g_{\la,\mu}$. 

By direct computation, one may verify that the three functions in the denominators of \eqref{functions} have no zero values inside the fundamental region $F$ of $C_2$. Moreover, whenever one of the denominator functions turns out to be zero at the boundary $\partial F$, the family of corresponding numerator functions is also zero there. More precisely, the functions $c^-$ are zero at the long side of the triangle $F$, and the functions $s^+$ are zero at the short sides of $F$. Such behaviour is known in general only for $S_\rho$, where $\rho$ is the half sum of positive roots of any compact simple $G$. In our case, $-4\sin^-_{(2,1)}(x,y)=S^-_{(1,1)}(x,y)=S_{(1,1)}(x,y)$.
 
Our main tool in constructing the polynomials is the decomposition of products of pairs of functions \eqref{trigfunctions}, besides the choice of variables. Since all products were decomposed in \cite{HMP}, we refrain here from describing methods of finding the decompositions.

Following the general idea of \cite{NPT}, the lowest functions $c^+$ multiplied by $2$ serve as the polynomial variables $X$ and $Y$ for all four cases \eqref{functions}:
\begin{equation}\label{variables}
X:=2c^+_{(1,0)}(x,y)=2(\cos(\pi x)+\cos(\pi y))\,,\quad 
Y:=2c^+_{(1,1)}(x,y)=4\cos(\pi x)\cos(\pi y)\,, 
\quad x,y\in\R\,.
\end{equation}
Clearly we also have $c^+_{(0,0)}(x,y)=2$. In addition, we need the three variables appearing as denominator terms in \eqref{functions}:
\begin{equation}\label{UVW}
\begin{alignedat}{3}
U &:=2c^-_{(1,0)}(x,y)\,,&\qquad 
V &:=2s^+_{(1,1)}(x,y)\,,&\qquad
W &:=4s^-_{(2,1)}(x,y)\,.\\
&\# U=1\,,&\qquad
&\# V=0\,,&\qquad
&\# W=1\,,&
\end{alignedat}
\end{equation}

\subsection{The polynomials $\cos^+$}\ 

According to Table~\ref{products}, polynomials of type $c^+$ arise in decompositions of products of four different types, namely $\cos^+\cos^+$, $\cos^-\cos^-$, $\sin^+\sin^+$, and $\sin^-\sin^-$. Decompositions of products $X^2$ and $Y^2$ yield quadratic polynomials, both with the congruence number $\#=0$. The only polynomial quadratic in $X$ and $Y$ with $\#=1$ is $XY$. Calculating the products of two variables, we obtain
$$
2c^+_{(2,0)}=X^2-2Y-4\,,\qquad 2c^+_{(2,2)}=Y^2-2X^2+4Y+4
$$
Taking ever higher functions $c^\pm_{(\lambda,\mu)}$, multiplying them by either $X$ or $Y$, and decomposing the products, we obtain higher and higher degree polynomials. Continuing with polynomial of $\#=0$, we obtain the next lowest polynomials,
\begin{equation}
\begin{aligned}
&4c^+_{(3,1)}=X^2Y-2Y^2-6Y\\
&2c^+_{(3,3)}=Y^3-3X^2Y+6Y^2+9Y\\
&2c^+_{(4,0)}=X^4-4X^2Y+2Y^2-4X^2+8Y+4\\
&4c^+_{(4,2)}=X^2Y^2-2X^4-2Y^3+8X^2Y-12Y^2+10X^2-20Y-8\\
&2c^+_{(4,4)}=Y^4-4X^2Y^2+2X^4+8Y^3-8X^2Y+20Y^2-8X^2+16Y+4
\end{aligned}
\end{equation}

See Table~\ref{cps} for more $c^+$ polynomials with $\#=0$.

Decompositions of the products with $\#=1$ are obtained in a similar way:
\begin{equation}
\begin{aligned}
&4c^+_{(2,1)}=XY-2X \\
&2c^+_{(3,0)}=X^3-3XY-3X\\
&4c^+_{(3,2)}=XY^2-2X^3+3XY+6X \\
&4c^+_{(4,1)}=X^3Y-3XY^2-4XY+2X \\
&4c^+_{(4,3)}=XY^3-3X^3Y+5XY^2+2X^3+6XY-6X \\
\end{aligned}
\end{equation}
Note that there is an easy useful check on any recurrence relation or decomposition of a product. Indeed, specializing $(x,y)=(0,0)$, we get  
\begin{equation}\label{check}
c_{(\lambda,\mu)}^+(0,0)=2\,,\qquad X(0,0)=Y(0,0)=4\,,
\end{equation} 
which turns any recurrence relation into an equality of integers.

See Table~\ref{cps} and Table~\ref{cpl} for more polynomials with $\#=0$ and $1$ respectively.

\subsection{Recurrence relations for symmetric cosine functions $c^+_{(\la,\mu)}$}\ 
\label{c+recurrence}

Using the variables \eqref{variables} as before, and decomposing appropriate products of $c^+$ functions, we have
\begin{equation}\label{c+recurs}
\begin{alignedat}{2}
\la\geq2,\,\mu=0 &:&\quad 2c^+_{(\la,0)}
      &=4c^+_{(\la-1,0)} c^+_{(1,0)}-2c^+_{(\la-2,0)}-4c^+_{(\la-1,1)}\\
\la-1>\mu>0      &:&\quad 4c^+_{(\la,\mu)}
      &=8c^+_{(\la-1,\mu)}c^+_{(1,0)}-4c^+_{(\la-2,\mu)}-4c^
                       +_{(\la-1,\mu+1)}-4c^+_{(\la-1,\mu-1)}\\
\la-1=\mu>0      &:&\qquad 4c^+_{(\la,\la-1)}
      &=4c^+_{(\la-1,\la-1)} c^+_{(1,0)}-4c^+_{(\la-1,\la-2)}\\
\la=\mu\geq2 &:&\quad 2c^+_{(\la,\la)}
      &=4c^+_{(\la-1,\la-1)} c^+_{(1,1)}-4c^+_{(\la,\la-2)}-2c^+_{(\la-2,\la-2)}
\end{alignedat}
\end{equation}
Again the check \eqref{check} is valid in \eqref{c+recurs}.

Starting from the lowest admissible values of $\la$ and $\mu$, we recursively build $c^+$ polynomials up to any desired degree in terms of variables $X$ and $Y$ using \eqref{c+recurs}. Some insight into the structure of recurrence relations is gained when they are rewritten as 3-term relations with matrix coefficients \cite{kow1,kow2}.

Introduce the polynomials renormalized by the factor \eqref{normalization}
$P_{\la,\mu}:=g_{\la\mu}c^+_{(\la,\mu)}$. Form the vector $\mathbb{P}_\la$,
\begin{gather}
\mathbb{P}_\la:=\left(P_{\la,0}(X,Y),P_{\la,1}(X,Y),\dots,P_{\la,\la-1}(X,Y),P_{\la,\la}(X,Y)\right)^T\in\R^{\la+1,1}
\end{gather}
and define on $\mathbb{P}_\la$  the operation
\begin{gather}
(X,Y)\mathbb{P}_\la:=\left(X\mathbb{P}_\la^T,Y\mathbb{P}_\la^T\right)^T\in\R^{2\la+2,1}
\end{gather}
Assuming that the lowest polynomials with $\la\leq3$ have been calculated, see Table~\ref{cps} and Table~\ref{cpl}, the polynomials with  $\la+1\geq4$ are found from the 3-term recurrence relation:
\begin{equation}
\mathbb{P}_{\la+1}=\mathbb{D}_\la(X,Y)\mathbb{P}_\la+\mathbb{E}_\la\mathbb{P}_\la+\mathbb{F}_\la\mathbb{P}_{\la-1}\,,
\end{equation}
where the matrix coefficients are the following,
$$
\mathbb{D}_\la=\begin{pmatrix} 1\\
                                 &\ddots\\
                                 &&1\\
                                 &-1&0&&&&1 \end{pmatrix}\in\R^{\la+2,2\la+2}\,,
$$

$$
\mathbb{F}_\la=\begin{pmatrix}-1\\
                                &\ddots \\
                                &&-1\\
                                &&&-2\\
                                &&&0\\
                                &&&1\end{pmatrix}\in\R^{\la+2,\la}\,,
\qquad
\mathbb{E}_\la=\begin{pmatrix}0&-1&\\
                                -2&0&-1&\\
                                &-1&0&-1&\\
                                &&\ddots&\ddots&\ddots\\                                 
                                &&&-1&0&-1&\\
                                &&&&-1&0&-2\\
                                &&&&0&-1&0\\
                                &&&&1&0&2 \end{pmatrix}\in\R^{\la+2,\la+1}\,.
$$

\subsection{Recurrence relations for antisymmetric cosine functions $c^-_{(\la,\mu)}$}\ 
\label{c-recurrence}

Polynomials of type $c^-$ arise in decompositions of two kinds of products \eqref{products}, namely $\cos^+\cos^-$, and $\sin^+\sin^-$. Recurrence relations for the function $c^-$ are described in terms of the variables $X$, $Y$, and $U$, where $\tfrac12U=c^-_{(1,0)}$ is the lowest of the $c^-$-functions. 

There are two second degree polynomials
$$
2c^-_{(2,0)}=UX\,,\qquad 4c^-_{(2,1)}=U(Y+2)\,.
$$
Decompositions of appropriate products yield the recurrence relations for higher degree polynomials.
\begin{equation}\label{c-recurs}
\begin{alignedat}{2}
\la>2,\,\mu=0 &:&\quad 2c^-_{(\la,0)}
    &=4c^-_{(\la-1,0)}c^+_{(1,0)}-2c^-_{(\la-2,0)}-4c^-_{(\la-1,1)}\\
\la-2>\mu>0          &:& \quad 4c^-_{(\la,\mu)}
    &=8c^-_{(\la-1,\mu)}c^+_{(1,0)}-4c^-_{(\la-2,\mu)}
    -4c^-_{(\la-1,\mu+1)}-4c^-_{(\la-1,\mu-1)}\\
\la> 2,\,\la-1=\mu>0 &:& \quad 4c^-_{(\la,\la-1)}
    &=8c^-_{(\la-1,\la-2)} c^+_{(1,1)}-4c^-_{(\la,\la-3)}+4c^-_{(\la-1,\la-2)}-4c^-_{(\la-2,\la-3)}\\
\la-2=\mu>0   &:&\quad 4c^-_{(\la,\la-2)}
    &=8c^-_{(\la-1,\la-2)} c^+_{(1,0)}-4c^-_{(\la-1,\la-3)}
      \end{alignedat}
\end{equation}
 Using \eqref{c-recurs} and starting from the lowest admissible values of $\la$ and $\mu$, we build the $c^-$ polynomials up to any desired degree in $X$, $Y$, and $U$, where $U$ occurs in the first power only. Instead it is more interesting to construct the polynomials $c^-_{(\la,\mu)}/U$ in order to compare them with the other polynomials of the functions \eqref{functions}, all in $X$, $Y$ variables.
 
Recurrence relations can be written in their 3-term form for the polynomials and solved provided the polynomials  $c^-_{(\la,\mu)}/U$ for $\la\leq4$ are known. See Table~\ref{cms} and Table~\ref{cml}.
\begin{gather}
P_{\la,\mu}=g_{\la,\mu}c^-_{(\la,\mu)}/U\,.
\end{gather} 
Introducing the vector $\mathbb{P}_\la$ with polynomial components,
\begin{gather}
\mathbb{P}_\la:=\left(P_{\la,0}(X,Y),P_{\la,1}(X,Y),\dots,P_{\la,\la-2}(X,Y),P_{\la,\la-1}(X,Y)\right)^T\in\R^{\la,1}
\end{gather}
and the transformation
\begin{gather}
(X,Y)\mathbb{P}_\la :=\left(X\mathbb{P}_\la^T,Y\mathbb{P}_\la^T\right)^T\in\R^{2\la,1}
\end{gather}
we get the recurrence relations for $\la+1\geq5$
\begin{equation}
\mathbb{P}_{\la+1}=\mathbb{D}_\la(X,Y)\mathbb{P}_\la+\mathbb{E}_\la\mathbb{P}_\la+\mathbb{F}_\la\mathbb{P}_{\la-1}
\end{equation}
where
$$
\mathbb{D}_\la=\begin{pmatrix} 1\\
                                 &\ddots\\
                                 &&1\\
                                 &-1&0&&&&1 \end{pmatrix}\in\R^{\la+1,2\la}\,,
                                 \qquad \qquad
\mathbb{F}_\la=\begin{pmatrix}-1\\
                                &\ddots \\
                                &&-1\\
                                &&0\\
                                &&0\end{pmatrix}\in\R^{\la+1,\la-1}\,,
$$

$$\mathbb{E}_\la=\begin{pmatrix}0&-1&\\
                                -2&0&-1&\\
                                &-1&0&-1&\\
                                &&\ddots&\ddots&\ddots\\                                 
                                &&&-1&0&-1&\\
                                &&&&-1&0&-1\\
                                &&&&0&-1&0\\
                                &&&&1&0&2 \end{pmatrix}\in\R^{\la+1,\la}\,.
$$

\subsection{Recurrence relations for symmetric sine functions $s^+_{(\la,\mu)}$}\ 
\label{s+recurrence}

Polynomials of type $s^+$ arise in decompositions of two kinds of products \eqref{products}, namely $\cos^+\sin^+$, and $\cos^-\sin^-$. Recurrence relations are constructed for polynomials $s^+_{(\la,\mu)}$ in variables $X$, $Y$, and $V$, where $\tfrac12V=s^+_{(1,1)}$ is the lowest of the $s^+$-functions. 

The quadratic polynomials being $4s^+_{(2,1)}=VX$ and $2s^+_{(2,2)}=VY$, the higher degree polynomials are found from the relations
\begin{equation}
\begin{alignedat}{2}
\la> 2,\,\mu=1 &:&\quad 4s^+_{(\la,1)}
   &=8s^+_{(\la-1,1)}c^+_{(1,0)}-4s^+_{(\la-2,1)}-4s^+_{(\la-1,2)}\\
\la-1>\mu>1    &:&\quad 4s^+_{(\la,\mu)}
   &=8s^+_{(\la-1,\mu)}c^+_{(1,0)}-4s^+_{(\la-2,\mu)}-4s^+_{(\la-1,\mu+1)}
        -4s^+_{(\la-1,\mu-1)}\\
\la-1=\mu>1    &:&\quad 4s^+_{(\la,\la-1)}
   &=4s^+_{(\la-1,\la-1)} c^+_{(1,0)}-4s^+_{(\la-1,\la-2)}\\
\la=\mu>2&:&\quad 2s^+_{(\la,\la)}
   &=4s^+_{(\la-1,\la-1)} c^+_{(1,1)}-4s^+_{(\la,\la-2)}-2s^+_{(\la-2,\la-2)}
\end{alignedat}
\end{equation}
As in the previous subsection, we are more interested in polynomials $s^+/V$, which depend on $X$ and $Y$ only and therefore can be compared with polynomials of types \eqref{functions}.

See Table~\ref{sps} and Table~\ref{spl} of polynomials $s^+_{(\la,\mu)}/V$.

Putting
$$
P_{\la,\mu}=g_{\la,\mu}s^+_{(\la,\mu)}/V
$$
and assuming that the polynomials are known for  $\la\leq3$, see  Table~\ref{sps} and Table~\ref{spl}, we introduce the vector $\mathbb{P}_\la$ as before,
\begin{gather}
\mathbb{P}_\la:=\left(P_{\la,1}(X,Y),P_{\la,2}(X,Y),\dots,P_{\la,\la-1}(X,Y),P_{\la,\la}(X,Y)\right)^T\in\R^{\la,1}
\end{gather}
together with the transformation
\begin{gather}
(X,Y)\mathbb{P}_\la:=\left(X\mathbb{P}_\la^T,Y\mathbb{P}_\la^T\right)^T\in\R^{2\la,1}\,,
\end{gather}
we have the reccurence relations for $\la+1\geq4$:
\begin{equation}
\mathbb{P}_{\la+1}=\mathbb{D}_\la(X,Y)\mathbb{P}_\la+\mathbb{E}_\la\mathbb{P}_\la+\mathbb{F}_\la\mathbb{P}_{\la-1}\,.
\end{equation}
Here
$$
\mathbb{D}_\la=\begin{pmatrix} 1\\
                                 &\ddots\\
                                 &&1\\
                                 &-1&0&&&&1 \end{pmatrix}\in\R^{\la+1,2\la}\,,
                                 \qquad \qquad
\mathbb{F}_\la=\begin{pmatrix}-1\\
                                &\ddots \\
                                &&-1\\
                                &&&-2\\
                                &&&0\\
                                &&&1\end{pmatrix}\in\R^{\la+1,\la-1}\,,
$$

$$
\mathbb{E}_\la=\begin{pmatrix}0&-1&\\
                                -1&0&-1&\\
                                &\ddots&\ddots&\ddots\\                                 
                                &&-1&0&-1&\\
                                &&&-1&0&-2\\
                                &&&0&-1&0\\
                                &&&1&0&2 \end{pmatrix}\in\R^{\la+1,\la}\,.
$$

\subsection{Recurrence relations for antisymmetric sine functions $s^-_{(\la,\mu)}$}\
\label{s-recurrence}

Polynomials of type $s^-$ arise in decompositions of two kinds of products \eqref{products}, namely $\cos^+\sin^-$, and $\cos^-\sin^+$. The variables in this case are $X$, $Y$, and $W$, where $\frac14W=s^-_{(2,1)}$ is the lowest of $s^-$-functions. 

The two quadratic polynomials $4s^-_{(3,1)}=WX$ and $4s^-_{(3,2)}=W(Y+1)$ are then used together with the relations
\begin{equation}
\begin{alignedat}{2}
\la> 3,\, \mu=1 &:&\quad 
   4s^-_{(\la,1)}&=8s^-_{(\la-1,1)} c^+_{(1,0)}-4s^-_{(\la-2,1)}-4s^-_{(\la-1,2)}\\
\la-2>\mu>1     &:&\quad
   4s^-_{(\la,\mu)}&=8s^-_{(\la-1,\mu)}c^+_{(1,0)}-4s^-_{(\la-2,\mu)}
       -4s^-_{(\la-1,\mu+1)}-4s^-_{(\la-1,\mu-1)}\\
\la-2=\mu>1     &:&\quad 
   4s^-_{(\la,\la-2)}&=8s^-_{(\la-1,\la-2)} c^+_{(1,0)}-4s^-_{(\la-1,\la-3)}\\
\la> 3,\, \la-1=\mu>1 &:&\quad 
   4s^-_{(\la,\la-1)}&=8s^-_{(\la-1,\la-2)} c^+_{(1,1)}
       +4s^-_{(\la-1,\la-2)}-4s^-_{(\la,\la-3)}-4s^-_{(\la-2,\la-3)}
\end{alignedat}
\end{equation}
As in the previous subsection, we are interested in polynomials $s^-/W$ which depend on $X$ and $Y$ only and therefore can be compared with polynomials of types \eqref{functions}.

See Table~\ref{sms} and Table~\ref{sml} of polynomials $s^-_{(\la,\mu)}/W$.

In the case  $s^-_{(\la,\mu)}/W$, the polynomials 
$$
P_{\la,\mu}=g_{\la,\mu}s^-_{(\la,\mu)}/W
$$
are suitably normalized irreducible characters of the simple Lie group $C_2$. Introducing the vector $\mathbb{P}_\la$
\begin{equation}
\mathbb{P}_\la:=\left(P_{\la,1}(X,Y),P_{\la,2}(X,Y),\dots,P_{\la,\la-2}(X,Y),P_{\la,\la-1}(X,Y)\right)^T\in\R^{\la-1,1}
\end{equation}
and the transformation
\begin{equation}
(X,Y)\mathbb{P}_\la:=\left(X\mathbb{P}_\la^T,Y\mathbb{P}_\la^T\right)^T\in\R^{2\la-2,1}\,,
\end{equation}
we get the recurrence relations for $\la\geq4$:
\begin{equation}
\mathbb{P}_{\la+1}=\mathbb{D}_\la(X,Y)\mathbb{P}_\la+\mathbb{E}_\la\mathbb{P}_\la+\mathbb{F}_\la\mathbb{P}_{\la-1}\,,
\end{equation}
where
$$
\mathbb{D}_\la=\begin{pmatrix} 1\\
                                 &\ddots\\
                                 &&1\\
                                 &-1&0&&&&1 \end{pmatrix}\in\R^{\la,2\la-2}\,,\qquad \qquad
                                 \mathbb{F}_\la=\begin{pmatrix}-1\\
                                &\ddots \\
                                &&-1\\
                                &&0\\
                                &&0\end{pmatrix}\in\R^{\la,\la-2}\,,
$$

$$
\mathbb{E}_\la=\begin{pmatrix}0&-1&\\
                                -1&0&-1&\\
                                &-1&0&-1&\\
                                &&\ddots&\ddots&\ddots\\                                 
                                &&&-1&0&-1\\
                                &&&0&-1&0\\
                                &&&1&0&2 \end{pmatrix}\in\R^{\la,\la-1}\,.
$$

\section{Laplace operators}\ 
The functions \eqref{trigfunctions} are eigenfunctions of the Laplace operator $\Delta$ in variables $x$ and $y$ given relative to an orthogonal basis. The functions are also eigenfunctions of an algebraically independent degree four operator, say $\Gamma$, which is easily verified by direct computation.  
$$
\Delta=\frac{\partial^2}{\partial x^2}+\frac{\partial^2}{\partial y^2}\,,
\qquad
\Gamma =\frac{\partial^2}{\partial x^2}\frac{\partial^2}{\partial y^2}\,.
$$

After the substitution of variables \eqref{variables} and \eqref{UVW}, the operators $\Delta$ and $\Gamma$ are substantially transformed. When the operators are applied to the functions \eqref{trigfunctions} expressed in the variables $X,Y$ and $U,V,W$ respectively, we obtain two operators for each family of polynomials \eqref{functions}, denoted here by $\overline{\Delta},\overline{\Gamma}$ with following property: The polynomials corresponding to one of our four family of functions \eqref{functions} are eigenfunctions of corresponding $\overline{\Delta},\overline{\Gamma}$ \cite{Koorn}. There are minute differences between them. Rather than explain the differences, we write the two operators for all four cases. Note that the eigenvalues of $\overline{\Delta}$ are zero for the lowest values of $\la$ and $\mu$, admissible for each family of polynomials by the inequalities \eqref{inequalities}.

\subsection{The operators $\overline{\Delta}$ and $\overline{\Gamma}$ with eigenfunctions $\cos^+_{(\la,\mu)}$}\ 

Any polynomial $P_{\la,\mu}(X,Y)$ of the $\cos^+_{(\la,\mu)}$ family in variables \eqref{variables} is the eigenfunction of the operator $\overline{\Delta}=\frac1{\pi^2}\Delta$.
\begin{gather}
\overline{\Delta} P_{\la,\mu}(X,Y)=-(\la^2+\mu^2)P_{\la,\mu}(X,Y)\,,
\end{gather}
where
\begin{gather}
\overline{\Delta} P_{\la,\mu}(X,Y)= \left((8-X^2+2Y)\frac{\partial^2}{\partial X^2}
+(4X^2-8Y-2Y^2)\frac{\partial^2}{\partial Y^2}\right.\notag\\
\left.+2(4X-XY)\frac{\partial^2}{\partial X \partial Y}
-X\frac{\partial}{\partial X}
-2Y\frac{\partial}{\partial Y}\right) P_{\la,\mu}(X,Y)\,.
\end{gather}

Any polynomial $P_{\la,\mu}(X,Y)$ of the $\cos^+_{(\la,\mu)}$ family in variables \eqref{variables} is the eigenfunction of the fourth degree operator $\overline{\Gamma}=\frac1{\pi^4}\Gamma$, which is independent of the Laplace operator.

\begin{gather}
\overline{\Gamma} P_{\la,\mu}(X,Y)=\la^2\mu^2P_{\la,\mu}(X,Y)\,,
\end{gather}
where
\begin{gather}
\overline{\Gamma} P_{\la,\mu}(X,Y)=\left((16-4X^2+8Y+Y^2)\frac{\partial^4}{\partial X^4}+(32X-8X^3+16XY+2XY^2)\frac{\partial^4}{\partial X^3\partial Y}+\right.\notag\\
(32Y+16Y^2+2Y^3+16X^2+ X^2Y^2-4X^4)\frac{\partial^4}{\partial X^2\partial Y^2}+(32XY-8X^3Y+16XY^2+2XY^3)\frac{\partial^4}{\partial X\partial Y^3}+\notag\\
(16Y^2-4X^2Y^2+8Y^3+Y^4)\frac{\partial^4}{\partial Y^4}+(XY-4X)\frac{\partial^3}{\partial X^3}+(X^2Y-24X^2+40Y+6Y^2+64)
\frac{\partial^3}{\partial X^2\partial Y}+\notag\\
(-20X^3+36XY+7XY^2+64X)\frac{\partial^3}{\partial X\partial Y^2}+(64Y-20X^2Y+40Y^2+6Y^3) \frac{\partial^3}{\partial Y^3}+Y\frac{\partial^2}{\partial X^2}+\notag\\
\left.(3XY-8X)\frac{\partial^2}{\partial X\partial Y}+ (-16X^2+32Y+7Y^2+32)\frac{\partial^2}{\partial Y^2}+Y\frac{\partial}{\partial Y}\right)P_{\la,\mu}(X,Y)
\end{gather}

\subsection{The operators $\overline{\Delta}$ and $\overline{\Gamma}$ with eigenfunctions $\cos^-_{(\la,\mu)}/U$}\ 

Any polynomial $P_{\la,\mu}(X,Y)$ of the $\cos^-_{(\la,\mu)}/U$ family in variables \eqref{variables} is the eigenfunction of the operator $\overline{\Delta}=\frac1{\pi^2U}\left(\Delta+1\right)$
\begin{gather}
\overline{\Delta} \left(P_{\la,\mu}(X,Y)\right)=-(\la^2+\mu^2-1)P_{\la,\mu}(X,Y)\,,
\end{gather}
where
\begin{gather}
\overline{\Delta} \left(P_{\la,\mu}(X,Y)\right)=\left((8-X^2+2Y)\frac{\partial^2}{\partial X^2}
+(4X^2-8Y-2Y^2)\frac{\partial^2}{\partial Y^2}\right.\notag\\
\left.+2(4X-XY)\frac{\partial^2}{\partial X \partial Y}
-3X\frac{\partial}{\partial X}
-(4Y+8)\frac{\partial}{\partial Y}\right)P_{\la,\mu}(X,Y)
\end{gather}

Any polynomial $P_{\la,\mu}(X,Y)$ of the $\cos^-_{(\la,\mu)}/U$ family in variables \eqref{variables} is the eigenfunction of the fourth degree operator $\overline{\Gamma}=\frac{1}{\pi^4U}\Gamma$, algebraically independent of the Laplace operator.

\begin{gather}
\overline{\Gamma} P_{\la,\mu}(X,Y)=\la^2\mu^2P_{\la,\mu}(X,Y)\,,
\end{gather}
where

\begin{gather}
\overline{\Gamma} P_{\la,\mu}(X,Y)=\left((16-4X^2+8Y+Y^2)\frac{\partial^4}{\partial X^4}+(32X-8X^3+16XY+2XY^2)\frac{\partial^4}{\partial X^3\partial Y}+\right.\notag\\
(32Y+16Y^2+2Y^3+16X^2+ X^2Y^2-4X^4)\frac{\partial^4}{\partial X^2\partial Y^2}+(32XY-8X^3Y+16XY^2+2XY^3)\frac{\partial^4}{\partial X\partial Y^3}+\notag\\
(16Y^2-4X^2Y^2+8Y^3+Y^4)\frac{\partial^4}{\partial Y^4}+(XY-4X)\frac{\partial^3}{\partial X^3}+(X^2Y-32X^2+56Y+8Y^2+96)
\frac{\partial^3}{\partial X^2\partial Y}+\notag\\
(-28X^3+52XY+9XY^2+96X)\frac{\partial^3}{\partial X\partial Y^2}+(96Y-28X^2Y+56Y^2+8Y^3) \frac{\partial^3}{\partial Y^3}+(2Y+4)\frac{\partial^2}{\partial X^2}+\notag\\
\left.(5XY-8X)\frac{\partial^2}{\partial X\partial Y}+ (-36X^2+76Y+14Y^2+96)\frac{\partial^2}{\partial Y^2}+(4Y+8)\frac{\partial}{\partial Y}\right)P_{\la,\mu}(X,Y)
\end{gather}

\subsection{The operators $\overline{\Delta}$ and $\overline{\Gamma}$ with eigenfunctions $\sin^+_{(\la,\mu)}/V$}\ 

Any polynomial $P_{\la,\mu}(X,Y)$ of the $\sin^+_{(\la,\mu)}/V$ family in variables \eqref{variables} is the eigenfunction of the operator $\overline{\Delta}=\frac1{\pi^2V}\left(\Delta+2\right)$
\begin{gather}
\overline{\Delta} \left(P_{\la,\mu}(X,Y)\right)=-(\la^2+\mu^2-2)P_{\la,\mu}(X,Y)\,,
\end{gather}
where
\begin{gather}
\overline{\Delta} \left(P_{\la,\mu}(X,Y)\right)
=\left((8-X^2+2Y)\frac{\partial^2}{\partial X^2}
+(4X^2-8Y-2Y^2)\frac{\partial^2}{\partial Y^2}\right.\notag\\
\left.+2(4X-XY)\frac{\partial^2}{\partial X \partial Y}
-3X\frac{\partial}{\partial X}
-6Y\frac{\partial}{\partial Y}\right)P_{\la,\mu}(X,Y)
\end{gather}

Any polynomial $P_{\la,\mu}(X,Y)$ of the $\sin^+_{(\la,\mu)}/V$ family in variables \eqref{variables} is the eigenfunction of the fourth degree operator $\overline{\Gamma}=\frac{1}{\pi^4V}\left(\Gamma-1\right)$, algebraically independent of the Laplace operator.

\begin{gather}
\overline{\Gamma} P_{\la,\mu}(X,Y)=(\la^2\mu^2-1)P_{\la,\mu}(X,Y)\,,
\end{gather}
where

\begin{gather}
\overline{\Gamma} P_{\la,\mu}(X,Y)=\left((16-4X^2+8Y+Y^2)\frac{\partial^4}{\partial X^4}+(32X-8X^3+16XY+2XY^2)\frac{\partial^4}{\partial X^3\partial Y}+\right.\notag\\
(32Y+16Y^2+2Y^3+16X^2+ X^2Y^2-4X^4)\frac{\partial^4}{\partial X^2\partial Y^2}+(32XY-8X^3Y+16XY^2+2XY^3)\frac{\partial^4}{\partial X\partial Y^3}+\notag\\
(16Y^2-4X^2Y^2+8Y^3+Y^4)\frac{\partial^4}{\partial Y^4}+(3XY-12X)\frac{\partial^3}{\partial X^3}+(3X^2Y-40X^2+56Y+10Y^2+64)
\frac{\partial^3}{\partial X^2\partial Y}+\notag\\
(-28X^3+44XY+13XY^2+64X)\frac{\partial^3}{\partial X\partial Y^2}+(64Y-28X^2Y+56Y^2+10Y^3) \frac{\partial^3}{\partial Y^3}\notag\\
+(7Y+X^2-8)\frac{\partial^2}{\partial X^2}+(17XY-32X)\frac{\partial^2}{\partial X\partial Y}+ (-36X^2+72Y+25Y^2+32)\frac{\partial^2}{\partial Y^2}+3X\frac{\partial}{\partial X}\notag\\
\left.+15Y\frac{\partial}{\partial Y}\right)P_{\la,\mu}(X,Y)
\end{gather}

\subsection{The operators $\overline{\Delta}$ and $\overline{\Gamma}$ with eigenfunctions $\sin^-_{(\la,\mu)}/W$}\ 

Any polynomial $P_{\la,\mu}(X,Y)$ of the $\sin^-_{(\la,\mu)}/W$ family in variables \eqref{variables} is the eigenfunction of the operator $\overline{\Delta}=\frac1{\pi^2W}\left(\Delta+5\right)$
\begin{gather}
\overline{\Delta} \left(P_{\la,\mu}(X,Y)\right)=-(\la^2+\mu^2-5)P_{\la,\mu}(X,Y)\,,
\end{gather}
where
\begin{gather}
\overline{\Delta} \left(P_{\la,\mu}(X,Y)\right)
=\left((8-X^2+2Y)\frac{\partial^2}{\partial X^2}
+(4X^2-8Y-2Y^2)\frac{\partial^2}{\partial Y^2}\right.\notag\\
\left.+2(4X-XY)\frac{\partial^2}{\partial X \partial Y}
-5X\frac{\partial}{\partial X}
-(8Y+8)\frac{\partial}{\partial Y}\right)P_{\la,\mu}(X,Y)\,.
\end{gather}

Any polynomial $P_{\la,\mu}(X,Y)$ of the $\sin^-_{(\la,\mu)}/V$ family in variables \eqref{variables} is the eigenfunction of the degree four operator $\overline{\Gamma}=\frac1{\pi^4W}\left(\Gamma-4\right)$, algebraically independent of the Laplace operator.

\begin{gather}
\overline{\Gamma} P_{\la,\mu}(X,Y)=(\la^2\mu^2-4)P_{\la,\mu}(X,Y)\,,
\end{gather}
where

\begin{gather}
\overline{\Gamma} P_{\la,\mu}(X,Y)=\left((16-4X^2+8Y+Y^2)\frac{\partial^4}{\partial X^4}+(32X-8X^3+16XY+2XY^2)\frac{\partial^4}{\partial X^3\partial Y}+\right.\notag\\
(32Y+16Y^2+2Y^3+16X^2+ X^2Y^2-4X^4)\frac{\partial^4}{\partial X^2\partial Y^2}+(32XY-8X^3Y+16XY^2+2XY^3)\frac{\partial^4}{\partial X\partial Y^3}+\notag\\
(16Y^2-4X^2Y^2+8Y^3+Y^4)\frac{\partial^4}{\partial Y^4}+(3XY-12X)\frac{\partial^3}{\partial X^3}+(3X^2Y-48X^2+72Y+12Y^2+96)
\frac{\partial^3}{\partial X^2\partial Y}+\notag\\
(-36X^3+60XY+15XY^2+96X)\frac{\partial^3}{\partial X\partial Y^2}+(96Y-36X^2Y+72Y^2+12Y^3) \frac{\partial^3}{\partial Y^3}\notag\\
+(X^2+10Y+4)\frac{\partial^2}{\partial X^2}+(23XY-32X)\frac{\partial^2}{\partial X\partial Y}+ (-64X^2+140Y+38Y^2+96)\frac{\partial^2}{\partial Y^2}+5X\frac{\partial}{\partial X}\notag\\
\left.+(32Y+32)\frac{\partial}{\partial Y}\right)P_{\la,\mu}(X,Y)
\end{gather}

\section{Continuous orthogonality of the four families of polynomials}
Continuous orthogonality of functions within each of the families $\cos^\pm$, $\sin^\pm$ has been shown \cite{HP,HMP}. Recall that the functions of the same family are orthogonal when integrated over the triangle $F$ (see \eqref{triangle}). Our task here is to reformulate the orthogonality relations after the substitution of variable \eqref{variables}. The substitution transforms the domain of orthogonality integration $F$ into its image $\mathfrak F$ shown in Fig~\eqref{integrationdomains}. Practically one needs to calculate the Jacobian of the change of integration variables.

The Weyl formula \cite{BS} for the integration of class functions over a compact simple Lie group $G$ reduces the integration to the maximal torus $\mathbb T$ of $G$ and independently provides information about the Jacobian. Up to some normalization conventions, it is the function $S_\rho$ of $G$. Our functions are W-invariant class functions, therefore integration is reduced to the fundamental region of the affine Weyl group $W$ on $\mathbb T$. These are commonly drawn conclusions for the functions \eqref{C+} and \eqref{S-}. Corresponding conclusions for functions $C^-$ and $S^+$ of \eqref{C-} and \eqref{S+} are made only here.

\subsection{Orthogonality of the polynomials of $c^+$}\ 

Consider any $g_{\la\mu}c^+_{(\la,\mu)}(x,y)$ and $g_{\la'\mu'}c^+_{(\la',\mu')}(x,y)$ and their respective polynomials $P_{\la,\mu}$ and $P_{\la',\mu'}$.

Computing directly the absolute value of the Jacobian $\tfrac{D(X,Y)}{D(x,y)}$, we have 
\begin{align}\label{jacobian}
J&:=\abs{\tfrac{D(X,Y)}{D(x,y)}}=2\pi^2\sqrt{4s^-_{(2,1)}(x,y)s^-_{(2,1)}(x,y)}\notag\\
 &=2\pi^2\sqrt{c^+_{(4,2)}-c^+_{(4,0)}-c^+_{(2,0)}+2-c^+_{(3,3)}+2c^+_{(3,1)}-c^+_{(1,1)}}\notag\\
 &=2\pi^2\sqrt{\tfrac14(-4X^4+16X^2+24X^2Y-64Y-32Y^2-4Y^3+X^2Y^2)}\notag\\
  &=\pi^2\sqrt{(X^2-4Y)((Y+4)^2-4X^2)}=\pi^2\abs{UV}=\pi^2\abs{W}\,,
\end{align}
where $U$,  $V$, and $W$ are defined in \eqref{UVW}. The fact that the Jacobian can be factorized in this way is of importance in the sequel.

\begin{align}
&\int_0^1\int_y^1g_{\la\mu}c^+_{(\la,\mu)}g_{\la'\mu'}c^+_{(\la',\mu')}\,dx\, dy
	=\left(\int_{-4}^4\int_{-\frac{1}{2}Y-2}^{\frac{1}{2}Y+2}
	-\int_{\frac14X^2}^4\int_{-4}^4\right)J^{-1}P_{\la,\mu}(X,Y)P_{\la',\mu'}(X,Y)\,dX\,dY\notag\\ 
&\qquad=\frac{1}{\pi^2}\left(\int_{-4}^4\int_{-\frac{1}{2}Y-2}^{\frac{1}{2}Y+2}
-\int_{\frac{1}{4}X^2}^4\int_{-4}^4\right)
   \frac{P_{\la,\mu}(X,Y)P_{\la',\mu'}(X,Y)}{\sqrt{(X^2-4Y)((Y+4)^2-4X^2)}}\,dX\,dY
   =g_{\la\mu}\delta_{\la\la'}\delta_{\mu\mu'}\,.
\end{align}

\savebox{\Imagebox}{\includegraphics[width=5cm]{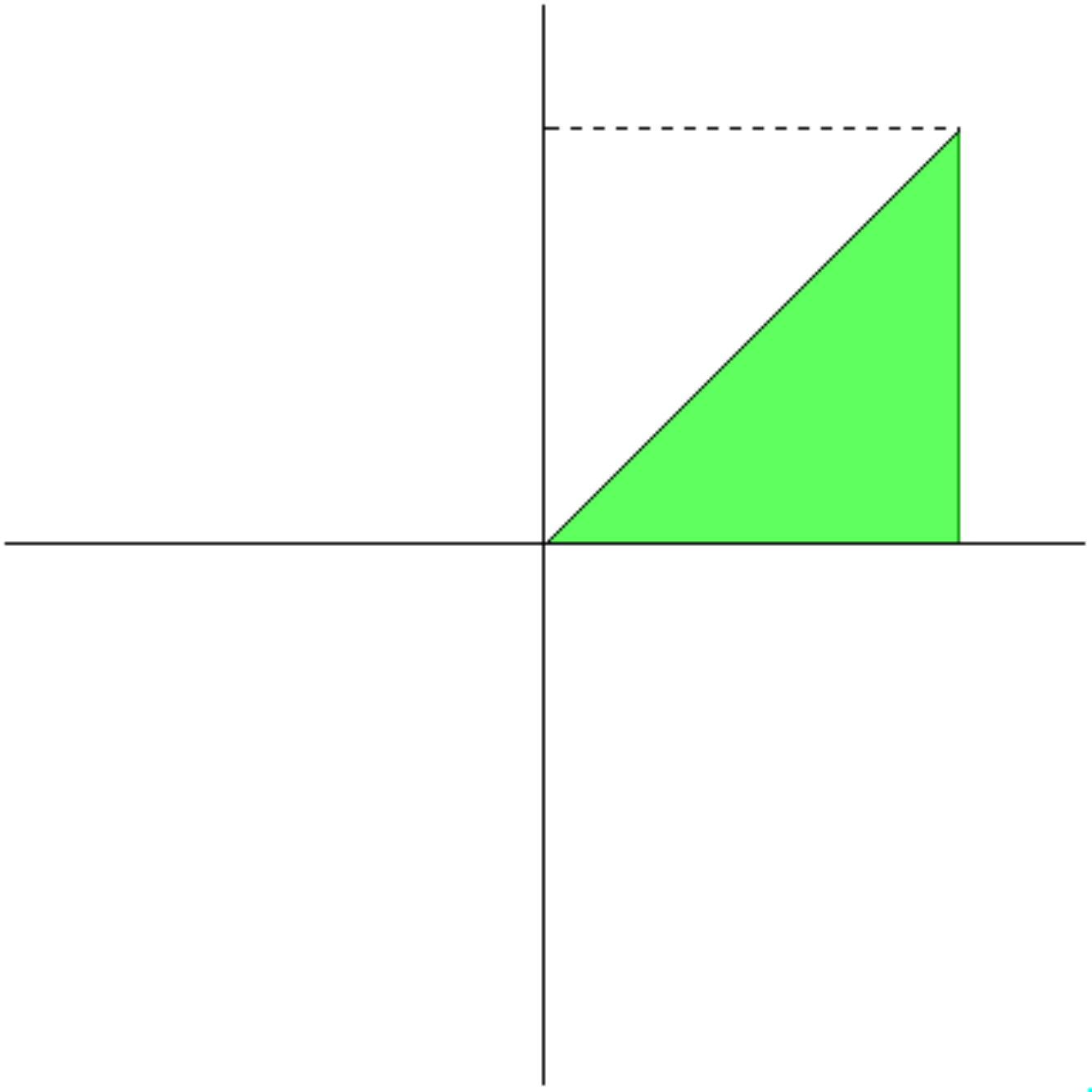}}

\begin{figure}[!ht]\label{integrationdomains}
\centering
\begin{pspicture}(\wd\Imagebox,\ht\Imagebox)
  \rput[lb](0,0){\usebox{\Imagebox}}
  \rput[t](2.38,2.45){\scriptsize{$0$}}
  \rput[t](4.35,2.45){\scriptsize{$1$}}
  \rput[t](4.9,2.4){\scriptsize{$x$}}
  \rput[t](2.4,4.38){\scriptsize{$1$}}
  \rput[t](2.38,4.95){\scriptsize{$y$}}
  \rput[t](3,3.65){\tiny{$y=x$}} 
\end{pspicture}
\savebox{\Imagebox}{\includegraphics[width=5cm]{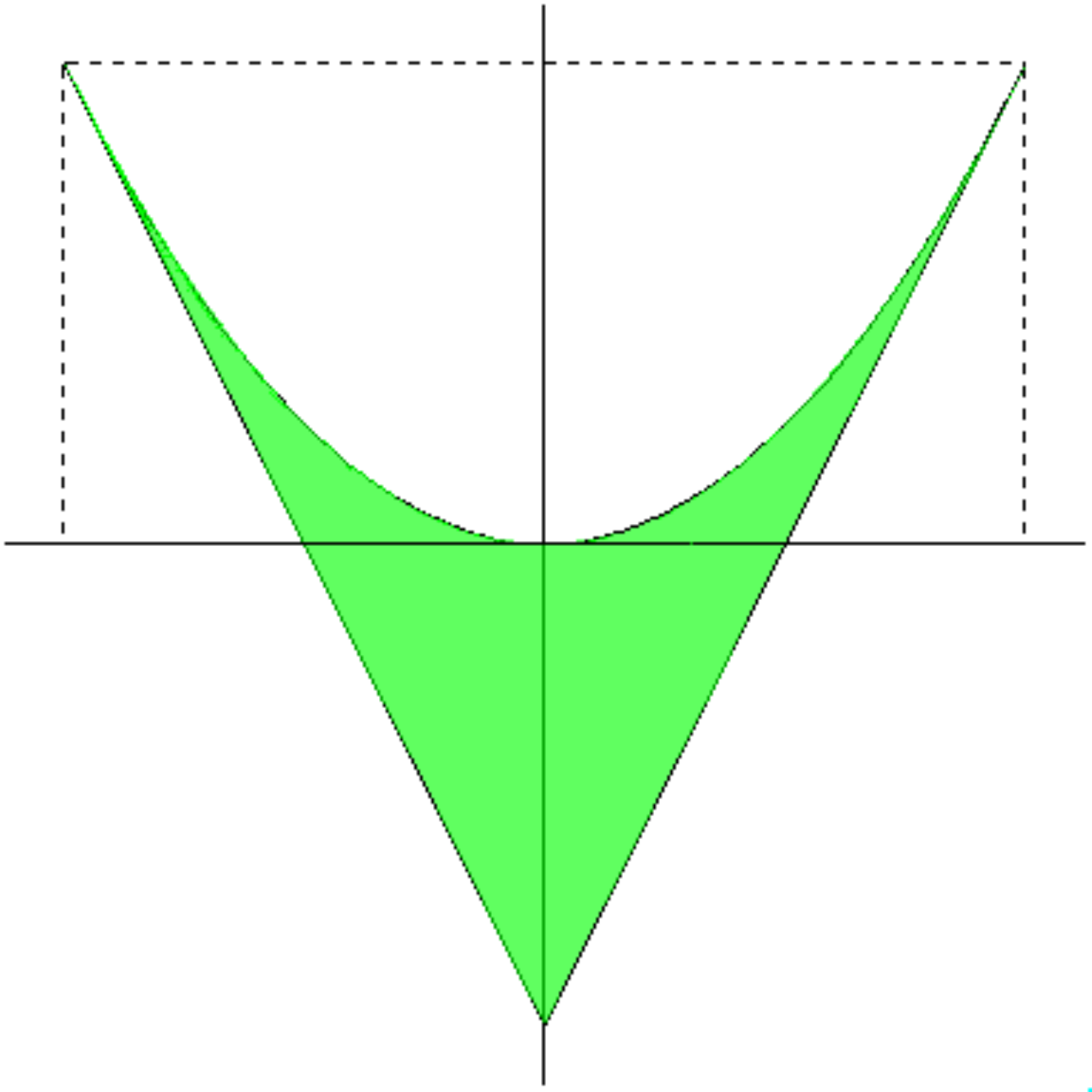}}
\hspace{2cm}
\begin{pspicture}(\wd\Imagebox,\ht\Imagebox)
  \rput[lb](0,0){\usebox{\Imagebox}}
  \rput[t](2.4,2.45){\scriptsize{$0$}}
  \rput[t](0.15,2.45){\scriptsize{$-4$}}
  \rput[t](4.65,2.45){\scriptsize{$4$}}
  \rput[t](4.95,2.45){\scriptsize{$X$}}
  \rput[t](2.4,4.6){\scriptsize{$4$}}
   \rput[t](2.28,0.3){\scriptsize{$-4$}}
  \rput[t](2.38,5){\scriptsize{$Y$}}
  \rput[t](3.15,3.5){\tiny{$Y=\frac14X^2$}}
   \rput[t](0.8,1.5){\tiny{$Y=-2(X+2)$}}
   \rput[t](4.1,1.5){\tiny{$Y=2(X-2)$}} 
\end{pspicture}
\caption{Fundamental domain $F$ in terms of the variables $x,y$, and its image $\mathfrak{F}$ after the change of variable \eqref{variables} to $X,\,Y$. The three equations defining the boundary of $\mathfrak{F}$ are shown.}
\end{figure}
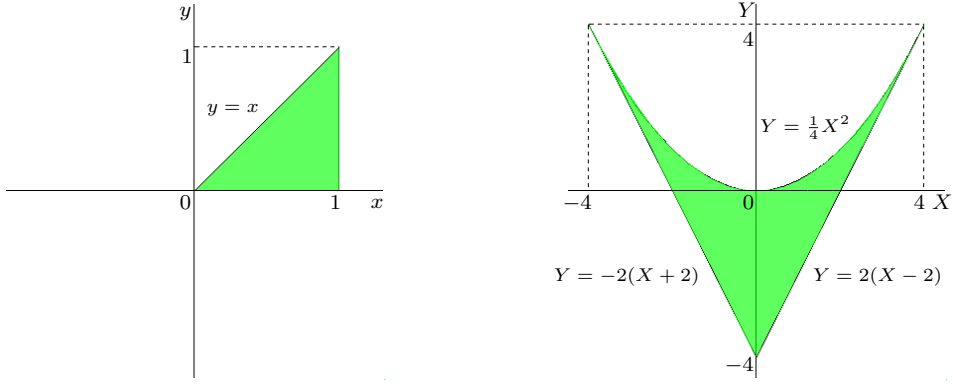

\subsection{Orthogonality of the polynomials of $c^-/U$}\ 

Consider any $g_{\la\mu}c^-_{(\la,\mu)}(x,y)$ and $g_{\la'\mu'}c^-_{(\la',\mu')}(x,y)$ and their respective polynomials $UP_{\la,\mu}$ and $UP_{\la',\mu'}$.

Multiplying $c^-_{(\la,\mu)}$ and $c^-_{(\la',\mu')}$, one needs to know 
$$
U^2=4c^-_{(1,0)}\cdot c^-_{(1,0)}=2c^+_{(2,0)}+4-4c^+_{(1,1)}=X^2-4Y\,,
$$
which is the first of the two factors of the Jacobian \eqref{jacobian}.
\begin{align}
&\int_0^1\int_y^1g_{\la\mu}c^-_{(\la,\mu)}g_{\la'\mu'}c^-_{(\la',\mu')}\,dx\,dy
	=\left(\int_{-4}^4\int_{-\frac12Y-2}^{\frac12Y+2}-\int_{\frac{1}{4}X^2}^4\int_{-4}^4\right)
   P_{\la,\mu}(X,Y)P_{\la',\mu'}(X,Y)U^2J^{-1}\,dX\,dY\notag\\ 
&\qquad=\frac1{\pi^2}\left(\int_{-4}^4\int_{-\frac12Y-2}^{\frac{1}{2}Y+2}-\int_{\frac{1}{4}X^2}^4\int_{-4}^4\right)
   P_{\la,\mu}(X,Y)P_{\la',\mu'}(X,Y)\sqrt{\frac{X^2-4Y}{(Y+4)^2-4X^2}}\,dX\,dY
   =g_{\la\mu}\delta_{\la\la'}\delta_{\mu\mu'}
\end{align}

\subsection{Orthogonality of the polynomials of $s^+/V$}\ 

Consider any $s^+_{(\la,\mu)}(x,y)$ and $s^+_{(\la',\mu')}(x,y)$ and their respective polynomials $VP_{\la\mu}$ and $VP{\la'\mu'}$. 

Multiplying $s^+_{(\la,\mu)}$ and $s^+_{(\la',\mu')}$, one needs to know 
$$
V^2=4s^+_{(1,1)}\cdot s^+_{(1,1)}=2c^+_{(2,2)}-4c^+_{(2,0)}+4=(Y+4)^2-4X^2\,,
$$
which is the second of the two factors of the Jacobian \eqref{jacobian}.
\begin{align}
&\int_0^1\int_y^1g_{\la\mu}s^+_{(\la,\mu)}g_{\la'\mu'}s^+_{(\la',\mu')}\,dx\,dy
	=\left(\int_{-4}^4\int_{-\frac12Y-2}^{\frac12Y+2}-\int_{\frac14X^2}^4
	\int_{-4}^4\right)P_{\la,\mu}(X,Y)P_{\la',\mu'}(X,Y)V^2J^{-1}\,dX\,dY\notag\\
&\q=\frac1{\pi^2}\left(\int_{-4}^4\int_{-\frac12Y-2}^{\frac12Y+2}
  -\int_{\frac14X^2}^4\int_{-4}^4\right)
   P_{\la,\mu}(X,Y)P_{\la',\mu'}(X,Y)\sqrt{\frac{(Y+4)^2-4X^2}{X^2-4Y}}\,dX\,dY=g_{\la\mu}\delta_{\la\la'}\delta_{\mu\mu'} 
\end{align}

\subsection{Orthogonality of the polynomials of the antisymmetric sine functions}\ 

Consider any $s^-_{(\la,\mu)}(x,y)$ and $s^-_{(\la',\mu')}(x,y)$ and their respective polynomials $WP_{\la\mu}$ and $WP_{\la'\mu'}$. 

Multiplying $s^-_{(\la,\mu)}$ and $s^-_{(\la',\mu')}$, one needs to know 
$$
W^2=4s^-_{(2,1)}\cdot s^-_{(2,1)}=\tfrac1{\pi^4}J^2\,,
$$
which is up to the constant $\pi^{-4}$ the square of the Jacobian \eqref{jacobian}.
\begin{align}
&\int_0^1\int_y^1g_{\la,\mu}s^-_{(\la,\mu)}g_{\la'\mu'}s^-_{(\la',\mu')}\,dx\,dy
	=\left(\int_{-4}^4\int_{-\frac{1}{2}Y-2}^{\frac{1}{2}Y+2}
	    -\int_{\frac{1}{4}X^2}^4\int_{-4}^4\right)
   P_{\la,\mu}(X,Y)P_{\la',\mu'}(X,Y)W^2J^{-1}\,dX\,dY\notag\\
&\qquad
   =\frac{1}{\pi^2}\left(\int_{-4}^4\int_{-\frac{1}{2}Y-2}^{\frac{1}{2}Y+2}-\int_{\frac{1}{4}X^2}^4\int_{-4}^4\right)
   P_{\la,\mu}(X,Y)P_{\la',\mu'}(X,Y)\sqrt{(X^2-4Y)((Y+4)^2-4X^2)}\,dX\,dY\notag\\
&\qquad=g_{\la\mu}\delta_{\la\la'}\delta_{\mu\mu'} 
\end{align}

\section{Identities of generalized trigonometric functions}\label{id}
Identities verified by the common trigonometric functions of one variable are numerous, well known, and often useful. Many more relations can be found for the two variable generalizations \eqref{trigfunctions} of the trigonometric functions. In this section, there are written relations for general values of parameters and the only condition on the parameters is that they are integers and distinct, i.e. they need not to satisfy the inequalities \eqref{inequalities}. Relations, obtained when some of the parameters coincide, are simpler. Numerous such relations are found in Appendix 1. Analogs of known identities of one variable trigonometric functions are more easily recognized there.

Relations are arranged according to the type of functions \eqref{trigfunctions} appearing on the right hand side of each equality.

\subsection{Generic relations for $c^+$}\ 
\begin{align}
&c^+_{(\la,\mu)}c^+_{(\la',\mu')}+c^-_{(\la,\mu)}c^-_{(\la',\mu')}+s^+_{(\la,\mu)}s^+_{(\la',\mu')}+s^-_{(\la,\mu)}s^-_{(\la',\mu')}=
    c^+_{(\la+\la',\mu+\mu')}+c^+_{(\la-\la',\mu-\mu')}\label{v1}\\
&c^+_{(\la,\mu)}c^+_{(\la',\mu')}+c^-_{(\la,\mu)}c^-_{(\la',\mu')}-s^+_{(\la,\mu)}s^+_{(\la',\mu')}-s^-_{(\la,\mu)}s^-_{(\la',\mu')}=
    c^+_{(\la+\la',\mu-\mu')}+c^+_{(\la-\la',\mu+\mu')}\\
&c^+_{(\la,\mu)}c^+_{(\la',\mu')}-c^-_{(\la,\mu)}c^-_{(\la',\mu')}+s^+_{(\la,\mu)}s^+_{(\la',\mu')}-s^-_{(\la,\mu)}s^-_{(\la',\mu')}=
    c^+_{(\la+\mu',\mu+\la')}+c^+_{(\la-\mu',\mu-\la')}\\
&c^+_{(\la,\mu)}c^+_{(\la',\mu')}-c^-_{(\la,\mu)}c^-_{(\la',\mu')}-s^+_{(\la,\mu)}s^+_{(\la',\mu')}+s^-_{(\la,\mu)}s^-_{(\la',\mu')}=
    c^+_{(\la+\mu',\mu-\la')}+c^+_{(\la-\mu',\mu+\la')}
\end{align}

\subsection{Generic relations for $c^-$}\
\begin{align}
&2\left(c^+_{(\la,\mu)}c^-_{(\la',\mu')}+s^+_{(\la,\mu)}s^-_{(\la',\mu')}\right)=
     c^-_{(\la+\la',\mu+\mu')}+c^-_{(\la-\la',\mu-\mu')}-c^-_{(\la+\mu',\mu+\la')}
     -c^-_{(\la-\mu',\mu-\la')}\\
&2\left(c^+_{(\la,\mu)}c^-_{(\la',\mu')}-s^+_{(\la,\mu)}s^-_{(\la',\mu')}\right)=
     c^-_{(\la+\la',\mu-\mu')}+c^-_{(\la-\la',\mu+\mu')}-c^-_{(\la+\mu',\mu-\la')}
     -c^-_{(\la-\mu',\mu+\la')}
\end{align}

\subsection{Generic relations for $s^+$}\
\begin{align}
&2\left(s^+_{(\la,\mu)}c^+_{(\la',\mu')}+s^-_{(\la,\mu)}c^-_{(\la',\mu')}\right)=
     s^+_{(\la+\la',\mu+\mu')}+s^+_{(\la+\la',\mu-\mu')}+s^+_{(\la-\la',\mu+\mu')}
     +s^+_{(\la-\la',\mu-\mu')}\\
&2\left(s^+_{(\la,\mu)}c^+_{(\la',\mu')}-s^-_{(\la,\mu)}c^-_{(\la',\mu')}\right)=
     s^+_{(\la+\mu',\mu+\la')}+s^+_{(\la+\mu',\mu-\la')}+s^+_{(\la-\mu',\mu+\la')}
     +s^+_{(\la-\mu',\mu-\la')}
\end{align}

\subsection{Generic relations for $s^-$}\ 
\begin{align}
&2\left(s^-_{(\la,\mu)}c^+_{(\la',\mu')}+s^+_{(\la,\mu)}c^-_{(\la',\mu')}\right)=
     s^-_{(\la+\la',\mu+\mu')}+s^-_{(\la+\la',\mu-\mu')}+s^-_{(\la-\la',\mu+\mu')}
     +s^-_{(\la-\la',\mu-\mu')}\\
&2\left(s^-_{(\la,\mu)}c^+_{(\la',\mu')}-s^+_{(\la,\mu)}c^-_{(\la',\mu')}\right)=
     s^-_{(\la+\mu',\mu+\la')}+s^-_{(\la+\mu',\mu-\la')}+s^-_{(\la-\mu',\mu+\la')}
     +s^-_{(\la-\mu',\mu-\la')}\label{v2}
\end{align}

\section{Conclusions}
The explicit relation between the functions studied here as `generalizations of 2-dimensional trigonometric functions' \eqref{trigfunctions}, and the same functions after the linear substitution of variables \eqref{substitution}, called the orbit functions of the simple Lie algebra or group of type $C_2$ (see \eqref{C+}--\eqref{S+}), has important practical and theoretical implications. Indeed, the functions are simpler in their `trigonometric' form \eqref{trigfunctions} than in their $C_2$ form, and are given in variables that refer to an orthogonal basis in $\R^2$. The Lie theoretical perspective on the functions allows one to discretize the orbit functions and carry over the discretization to the four families of $C_2$ polynomials.

It is easy to recognize that among the four families \eqref{functions} the last one is formed by the character of finite dimensional irreducible representations of $C_2$. Perhaps even more significant are the implications of the fact that the second and third families of functions in \eqref{functions} have not been encountered so far in the literature or in the Lie theory in general.

As a result of this paper, the number of families of orthogonal polynomials based on $C_2$ that can be built has grown from three ($C$-, $S$-, and $E$-functions of \cite{P}) to four ($C^\pm$ and $S^\pm$), and to another six $E$-like combinations $C^\pm+S^\pm$. 

Generalization of these facts to all simple Lie groups with two lengths of simple roots is imminent.

Interesting and undoubtedly useful are the new functions $C^-$ and $S^+$ due to their `hybrid' behaviour at the boundary of the region $F$, i.e. they are symmetric on some sides of the simplex $F$, skew-symmetric on the other sides.

There is another related but different point of view on the polynomials we constructed in this paper. Tables in Appendix~2 show examples of the polynomials in the form \eqref{functions}, whose variables $X$ and $Y$ are the two lowest $C$-functions \eqref{variables}. Alternatively, one may choose as the polynomial variables the characters 
$\chi_{\omega_1}$ and $\chi_{\omega_2}$ of the $C_2$ representations of the lowest dimension, namely 4 and 5. The character variables are then related to $X$ and $Y$ of \eqref{variables} as follows,
$$
\chi_{\omega_1}=X\,,\qquad \chi_{\omega_2}=Y+1\,.
$$
The characters have all of the properties required to build the polynomials, including continuous and discrete orthogonality \cite{MP87}. Often the characters are avoided in applications because, unlike orbit functions, they grow without limits for higher representations. However, on the level of polynomial exploitation, that difference is not visible. Hence the alternative can be rather useful \cite{MP2010}.

\section*{Acknowledgments}
We gratefully acknowledge the support of this work by the Natural Sciences and Engineering Research Council of Canada, by the MIND Research Institute of Santa Ana, California, and by the Doppler Institute of the Czech Technical University in Prague. LM is grateful for the hospitality extended to her at the Centre de recherches math\'ematiques, Universit\'e de Montr\'eal, where most of the work was done. LM is thankful for the support from the Czech Technical University grant SGS10/210/OHK4/2T/14. Both authors express their gratitude for the hospitality of the Doppler Institute, where the work was started.

\newpage
\section*{Appendix 1}
In addition to the generic recurrence relations of Sec.~\ref{id}, here we present simplified relations obtained when some of the parametres of \eqref{v1}--\eqref{v2} coincide.

\subsection*{Relations for $c^+$}

\begin{align*}
&\la'=\la\\
&c^+_{(\la,\mu)}c^+_{(\la,\mu')}+c^-_{(\la,\mu)}c^-_{(\la,\mu')}+s^+_{(\la,\mu)}s^+_{(\la,\mu')}+s^-_{(\la,\mu)}s^-_{(\la,\mu')}=
    c^+_{(2\la,\mu+\mu')}+c^+_{(0,\mu-\mu')}\\
&c^+_{(\la,\mu)}c^+_{(\la,\mu')}+c^-_{(\la,\mu)}c^-_{(\la,\mu')}-s^+_{(\la,\mu)}s^+_{(\la,\mu')}-s^-_{(\la,\mu)}s^-_{(\la,\mu')}=
    c^+_{(2\la,\mu-\mu')}+c^+_{(0,\mu+\mu')}\\
&c^+_{(\la,\mu)}c^+_{(\la,\mu')}-c^-_{(\la,\mu)}c^-_{(\la,\mu')}+s^+_{(\la,\mu)}s^+_{(\la,\mu')}-s^-_{(\la,\mu)}s^-_{(\la,\mu')}=
    c^+_{(\la+\mu',\mu+\la)}+c^+_{(\la-\mu',\mu-\la)}\\
&c^+_{(\la,\mu)}c^+_{(\la,\mu')}-c^-_{(\la,\mu)}c^-_{(\la,\mu')}-s^+_{(\la,\mu)}s^+_{(\la,\mu')}+s^-_{(\la,\mu)}s^-_{(\la,\mu')}=
    c^+_{(\la+\mu',\mu-\la)}+c^+_{(\la-\mu',\mu+\la)}\\
&\\
&\la'=\la,\q\mu'=\mu \\
&c^+_{(\la,\mu)}c^+_{(\la,\mu)}+c^-_{(\la,\mu)}c^-_{(\la,\mu)}+s^+_{(\la,\mu)}s^+_{(\la,\mu)}+s^-_{(\la,\mu)}s^-_{(\la,\mu)}=
    c^+_{(2\la,2\mu)}+2\\
&c^+_{(\la,\mu)}c^+_{(\la,\mu)}+c^-_{(\la,\mu)}c^-_{(\la,\mu)}-s^+_{(\la,\mu)}s^+_{(\la,\mu)}-s^-_{(\la,\mu)}s^-_{(\la,\mu)}=
    c^+_{(2\la,0)}+c^+_{(0,2\mu)}\\
&c^+_{(\la,\mu)}c^+_{(\la,\mu)}-c^-_{(\la,\mu)}c^-_{(\la,\mu)}+s^+_{(\la,\mu)}s^+_{(\la,\mu)}-s^-_{(\la,\mu)}s^-_{(\la,\mu)}=
    c^+_{(\la+\mu,\la+\mu)}+c^+_{(\mu-\la,\mu-\la)}\\
&c^+_{(\la,\mu)}c^+_{(\la,\mu)}-c^-_{(\la,\mu)}c^-_{(\la,\mu)}-s^+_{(\la,\mu)}s^+_{(\la,\mu)}+s^-_{(\la,\mu)}s^-_{(\la,\mu)}=
    2c^+_{(\la+\mu,\la-\mu)}\\
&\\
&\mu=\la\\
&c^+_{(\la,\la)}c^+_{(\la',\mu')}+s^+_{(\la,\la)}s^+_{(\la',\mu')}=
    c^+_{(\la+\la',\la+\mu')}+c^+_{(\la-\la',\la-\mu')}\\
&c^+_{(\la,\la)}c^+_{(\la',\mu')}-s^+_{(\la,\la)}s^+_{(\la',\mu')}=
    c^+_{(\la+\la',\la-\mu')}+c^+_{(\la-\la',\la+\mu')}\\
&\\
&\mu=\la,\q\mu'=\la'\\
&c^+_{(\la,\la)}c^+_{(\la',\la')}+s^+_{(\la,\la)}s^+_{(\la',\la')}=
    c^+_{(\la+\la',\la+\la')}+c^+_{(\la-\la',\la-\la')}\\
&c^+_{(\la,\la)}c^+_{(\la',\la')}-s^+_{(\la,\la)}s^+_{(\la',\la')}=
    2c^+_{(\la+\la',\la-\la')}\\
&\\
&\mu=\la'=\la\\
&c^+_{(\la,\la)}c^+_{(\la,\mu')}+s^+_{(\la,\la)}s^+_{(\la,\mu')}=
    c^+_{(2\la,\la+\mu')}+c^+_{(0,\la-\mu')}\\
&c^+_{(\la,\la)}c^+_{(\la,\mu')}-s^+_{(\la,\la)}s^+_{(\la,\mu')}=
    c^+_{(2\la,\la-\mu')}+c^+_{(0,\la+\mu')}\\
&\\
&\mu'=\la'=\mu=\la\\
&c^+_{(\la,\la)}c^+_{(\la,\la)}+s^+_{(\la,\la)}s^+_{(\la,\la)}=
    c^+_{(2\la,2\la)}+2\\
&c^+_{(\la,\la)}c^+_{(\la,\la)}-s^+_{(\la,\la)}s^+_{(\la,\la)}=
    2c^+_{(2\la,0)}\\
\end{align*}

\subsection*{Relations for $c^-$}

\begin{align*}
&\la'=\la\\
&2\left(c^+_{(\la,\mu)}c^-_{(\la,\mu')}+s^+_{(\la,\mu)}s^-_{(\la,\mu')}\right)=
     c^-_{(2\la,\mu+\mu')}-c^-_{(\la+\mu',\mu+\la)}
     -c^-_{(\la-\mu',\mu-\la)}\\
&2\left(c^+_{(\la,\mu)}c^-_{(\la,\mu')}-s^+_{(\la,\mu)}s^-_{(\la,\mu')}\right)=
     c^-_{(2\la,\mu-\mu')}-c^-_{(\la+\mu',\mu-\la)}
     -c^-_{(\la-\mu',\mu+\la)}\\
&\\
&\la'=\la,\q\mu'=\mu \\
&2\left(c^+_{(\la,\mu)}c^-_{(\la,\mu)}+s^+_{(\la,\mu)}s^-_{(\la,\mu)}\right)=
     c^-_{(2\la,2\mu)}\\
&2\left(c^+_{(\la,\mu)}c^-_{(\la,\mu)}-s^+_{(\la,\mu)}s^-_{(\la,\mu)}\right)=
     c^-_{(2\la,0)}+c^-_{(0,2\mu)}\\
&\\
&\mu=\la\\
&c^+_{(\la,\la)}c^-_{(\la',\mu')}+s^+_{(\la,\la)}s^-_{(\la',\mu')}=
     c^-_{(\la+\la',\la+\mu')}+c^-_{(\la-\la',\la-\mu')}\\
&c^+_{(\la,\la)}c^-_{(\la',\mu')}-s^+_{(\la,\la)}s^-_{(\la',\mu')}=
     c^-_{(\la+\la',\la-\mu')}+c^-_{(\la-\la',\la+\mu')}\\
&\\
&\mu=\la'=\la\\
&c^+_{(\la,\la)}c^-_{(\la,\mu')}+s^+_{(\la,\la)}s^-_{(\la,\mu')}=
     c^-_{(2\la,\la+\mu')}\\
&c^+_{(\la,\la)}c^-_{(\la,\mu')}-s^+_{(\la,\la)}s^-_{(\la,\mu')}=
     c^-_{(2\la,\la-\mu')}
\end{align*}

\subsection*{Relations for $s^+$}

\begin{align*}
&\la'=\la\\
&2\left(s^+_{(\la,\mu)}c^+_{(\la,\mu')}+s^-_{(\la,\mu)}c^-_{(\la,\mu')}\right)=
     s^+_{(2\la,\mu+\mu')}+s^+_{(2\la,\mu-\mu')}\\
&2\left(s^+_{(\la,\mu)}c^+_{(\la,\mu')}-s^-_{(\la,\mu)}c^-_{(\la,\mu')}\right)=
     s^+_{(\la+\mu',\mu+\la)}+s^+_{(\la+\mu',\mu-\la)}+s^+_{(\la-\mu',\mu+\la)}
     +s^+_{(\la-\mu',\mu-\la)}\\
&\\
&\la'=\la,\q\mu'=\mu \\
&2\left(s^+_{(\la,\mu)}c^+_{(\la,\mu)}+s^-_{(\la,\mu)}c^-_{(\la,\mu)}\right)=
     s^+_{(2\la,2\mu)}\\
&2\left(s^+_{(\la,\mu)}c^+_{(\la,\mu)}-s^-_{(\la,\mu)}c^-_{(\la,\mu)}\right)=
     s^+_{(\la+\mu,\la+\mu)}-s^+_{(\la-\mu,\la-\mu)}\\
&\\
&\mu=\la\\
&2s^+_{(\la,\la)}c^+_{(\la',\mu')}=
     s^+_{(\la+\la',\la+\mu')}+s^+_{(\la+\la',\la-\mu')}+s^+_{(\la-\la',\la+\mu')}
     +s^+_{(\la-\la',\la-\mu')}\\
&\\
&\mu=\la,\q\mu'=\la'\\
&2s^+_{(\la,\la)}c^+_{(\la',\la')}=
     s^+_{(\la+\la',\la+\la')}+2s^+_{(\la+\la',\la-\la')}+s^+_{(\la-\la',\la-\la')}\\
&\\
&\mu=\la'=\la\\
&2s^+_{(\la,\la)}c^+_{(\la,\mu')}=
     s^+_{(2\la,\la+\mu')}+s^+_{(2\la,\la-\mu')}\\
&\\
&\mu=\mu'=\la'=\la\\
&2s^+_{(\la,\la)}c^+_{(\la,\la)}=
     s^+_{(2\la,2\mu)}
\end{align*}

\subsection*{Relations for $s^-$}

\begin{align*}
&\la'=\la\\
&2\left(s^-_{(\la,\mu)}c^+_{(\la,\mu')}+s^+_{(\la,\mu)}c^-_{(\la,\mu')}\right)=
     s^-_{(2\la,\mu+\mu')}+s^-_{(2\la,\mu-\mu')}\\
&2\left(s^-_{(\la,\mu)}c^+_{(\la,\mu')}-s^+_{(\la,\mu)}c^-_{(\la,\mu')}\right)=
     s^-_{(\la+\mu',\mu+\la)}+s^-_{(\la+\mu',\mu-\la)}+s^-_{(\la-\mu',\mu+\la)}
     +s^-_{(\la-\mu',\mu-\la)}\\
&\\
&\la'=\la,\q\mu'=\mu \\
&2\left(s^-_{(\la,\mu)}c^+_{(\la,\mu)}+s^+_{(\la,\mu)}c^-_{(\la,\mu)}\right)=
     s^-_{(2\la,2\mu)}\\
&s^-_{(\la,\mu)}c^+_{(\la,\mu)}-s^+_{(\la,\mu)}c^-_{(\la,\mu)}=
     s^-_{(\la-\mu,\la+\mu)}\\
&\\
&\mu=\la\\
&2s^+_{(\la,\la)}c^-_{(\la',\mu')}=
     s^-_{(\la+\la',\la+\mu')}+s^-_{(\la+\la',\la-\mu')}+s^-_{(\la-\la',\la+\mu')}
     +s^-_{(\la-\la',\la-\mu')}\\
&\\
&\mu=\la'=\la\\
&2s^+_{(\la,\la)}c^-_{(\la,\mu')}=
     s^-_{(2\la,\la+\mu')}+s^-_{(2\la,\la-\mu')}
\end{align*}

\section*{Appendix 2}

There are two tables shown for each of the four function of \eqref{functions}, one for each congruence class \eqref{congruence}. The first table contains the functions of congruence class $\#=0$ and the second table contains the functions with $\#=1$.

\begin{table}[!h] 
\begin{center}
\small
\begin{tabular}{|c|@{\,}r@{\,}|@{\,}r@{\,}|@{\,}r@{\,}|@{\,}r@{\,}|@{\,}r@{\,}|@{\,}r@{\,}|@{\,}r@{\,}|@{\,}r@{\,}|@{\,}r@{\,}|@{\,}r@{\,}|@{\,}r@{\,}|@{\,}r@{\,}|@{\,}r@{\,}|@{\,}r@{\,}|@{\,}r@{\,}|@{\,}r@{\,}|@{\,}r@{\,}|@{\,}r@{\,}|@{\,}r@{\,}|@{\,}r@{\,}|}
\hline
$c^+$&\footnotesize{$1$}&\footnotesize{$Y$}&\footnotesize{$X^2$}&\footnotesize{$Y^2$}&\footnotesize{$X^2Y$}&\footnotesize{$Y^3$}&\footnotesize{$X^4$}&\footnotesize{$X^2Y^2$}&\footnotesize{$Y^4$}&\footnotesize{$X^4Y$}&\footnotesize{$X^2Y^3$}&\footnotesize{$Y^5$}&\footnotesize{$X^6$}&\footnotesize{$X^4Y^2$}&\footnotesize{$X^2Y^4$}&\footnotesize{$Y^6$}&\footnotesize{$X^6Y$}&\footnotesize{$X^4Y^3$}&\footnotesize{$X^2Y^5$}&\footnotesize{$Y^7$}\\ \hline
$\frac12(0,0)$&$1$&&&&&&&&&&&&&&&&&&&\\ \hline
$2(1,1)$&$0$&$1$&&&&&&&&&&&&&&&&&&\\ \hline
$2(2,0)$&$-4$&$-2$&$1$&&&&&&&&&&&&&&&&&\\ \hline
$2(2,2)$&$4$&$4$&$-2$&$1$&&&&&&&&&&&&&&&&\\ \hline
$4(3,1)$&$0$&$-6$&$0$&$-2$&$1$&&&&&&&&&&&&&&&\\ \hline
$2(3,3)$&$0$&$9$&$0$&$6$&$-3$&$1$&&&&&&&&&&&&&&\\ \hline
$2(4,0)$&$4$&$8$&$-4$&$2$&$-4$&$0$&$1$&&&&&&&&&&&&&\\ \hline
$4(4,2)$&$-8$&$-20$&$10$&$-12$&$8$&$-2$&$-2$&$1$&&&&&&&&&&&&\\ \hline
$2(4,4)$&$4$&$16$&$-8$&$20$&$-8$&$8$&$2$&$-4$&$1$&&&&&&&&&&&\\ \hline
$4(5,1)$&$0$&$10$&$0$&$10$&$-5$&$2$&$0$&$-4$&$0$&$1$&&&&&&&&&&\\ \hline
$4(5,3)$&$0$&$-30$&$0$&$-40$&$20$&$-16$&$0$&$12$&$-2$&$-3$&$1$&&&&&&&&&\\ \hline
$2(5,5)$&$0$&$25$&$0$&$50$&$-25$&$35$&$0$&$-20$&$10$&$5$&$-5$&$1$&&&&&&&&\\ \hline
$2(6,0)$&$-4$&$-18$&$9$&$-12$&$24$&$-2$&$-6$&$9$&$0$&$-6$&$0$&$0$&$1$&&&&&&&\\ \hline
$4(6,2)$&$8$&$40$&$-20$&$42$&$-48$&$16$&$12$&$-24$&$2$&$12$&$-4$&$0$&$-2$&$1$&&&&&&\\ \hline
$4(6,4)$&$-8$&$-52$&$26$&$-100$&$56$&$-70$&$-14$&$51$&$-20$&$-12$&$16$&$-2$&$2$&$-4$&$1$&&&&&\\ \hline
$2(6,6)$&$4$&$36$&$-18$&$105$&$-48$&$112$&$12$&$-72$&$54$&$12$&$-36$&$12$&$-2$&$9$&$-6$&$1$&&&&\\ \hline
$4(7,1)$&$0$&$-14$&$0$&$-28$&$14$&$-14$&$0$&$28$&$-2$&$-7$&$9$&$0$&$0$&$-6$&$0$&$0$&$1$&&&\\ \hline
$4(7,3)$&$0$&$42$&$0$&$98$&$-49$&$70$&$0$&$-84$&$20$&$21$&$-34$&$2$&$0$&$18$&$-4$&$0$&$-3$&$1$&&\\ \hline
$4(7,5)$&$0$&$-70$&$0$&$-210$&$105$&$-224$&$0$&$168$&$-108$&$-42$&$94$&$-24$&$0$&$-30$&$20$&$-2$&$5$&$-5$&$1$&\\ \hline
$2(7,7)$&$0$&$49$&$0$&$196$&$-98$&$294$&$0$&$-196$&$210$&$49$&$-161$&$77$&$0$&$42$&$-56$&$14$&$-7$&$14$&$-7$&$1$\\ \hline
\end{tabular}
\bigskip
\caption{The polynomials $c^+_{(\la,\mu)}$ of the congruence class $\#=0$ of degrees $\leq7$. In the first column the labels $(\la,\mu)$ are shown. The constants $\frac12$, 2 and 4 should be understood as multiplying the functions $c^+_{(\la,\mu)}$, making all the entries of the table integer.}
\label{cps}
\end{center}
\end{table}  

\begin{table}[!h] 
\begin{center}
\small
\begin{tabular}{|c|@{\,}r@{\,}|@{\,}r@{\,}|@{\,}r@{\,}|@{\,}r@{\,}|@{\,}r@{\,}|@{\,}r@{\,}|@{\,}r@{\,}|@{\,}r@{\,}|@{\,}r@{\,}|@{\,}r@{\,}|@{\,}r@{\,}|@{\,}r@{\,}|@{\,}r@{\,}|@{\,}r@{\,}|@{\,}r@{\,}|@{\,}r@{\,}|}
\hline
$c^+$&\footnotesize{$X$}&\footnotesize{$XY$}&\footnotesize{$X^3$}&\footnotesize{$XY^2$}&\footnotesize{$X^3Y$}&\footnotesize{$XY^3$}&\footnotesize{$X^5$}&\footnotesize{$X^3Y^2$}&\footnotesize{$XY^4$}&\footnotesize{$X^5Y$}&\footnotesize{$X^3Y^3$}&\footnotesize{$XY^5$}&\footnotesize{$X^7$}&\footnotesize{$X^5Y^2$}&\footnotesize{$X^3Y^4$}&\footnotesize{$XY^6$}\\ \hline
$2(1,0)$&$1$&&&&&&&&&&&&&&&\\ \hline
$4(2,1)$&$-2$&$1$&&&&&&&&&&&&&&\\ \hline
$2(3,0)$&$-3$&$-3$&$1$&&&&&&&&&&&&&\\ \hline
$4(3,2)$&$6$&$3$&$-2$&$1$&&&&&&&&&&&&\\ \hline
$4(4,1)$&$2$&$-4$&$0$&$-3$&$1$&&&&&&&&&&&\\ \hline
$4(4,3)$&$-6$&$6$&$2$&$5$&$-3$&$1$&&&&&&&&&&\\ \hline
$2(5,0)$&$5$&$15$&$-5$&$5$&$-5$&$0$&$1$&&&&&&&&&\\ \hline
$4(5,2)$&$-10$&$-25$&$10$&$-15$&$10$&$-3$&$-2$&$1$&&&&&&&&\\ \hline
$4(5,4)$&$10$&$10$&$-10$&$15$&$-5$&$7$&$2$&$-4$&$1$&&&&&&&\\ \hline
$4(6,1)$&$-2$&$9$&$0$&$18$&$-6$&$5$&$0$&$-5$&$0$&$1$&&&&&&\\ \hline
$4(6,3)$&$6$&$-21$&$-2$&$-45$&$18$&$-21$&$0$&$15$&$-3$&$-3$&$1$&&&&&\\ \hline
$4(6,5)$&$-10$&$15$&$10$&$35$&$-20$&$28$&$-2$&$-16$&$9$&$5$&$-5$&$1$&&&&\\ \hline
$2(7,0)$&$-7$&$-42$&$14$&$-35$&$35$&$-7$&$-7$&$14$&$0$&$-7$&$0$&$0$&$1$&&&\\ \hline
$4(7,2)$&$14$&$77$&$-28$&$84$&$-70$&$35$&$14$&$-35$&$5$&$14$&$-5$&$0$&$-2$&$1$&&\\ \hline
$4(7,4)$&$-14$&$-56$&$28$&$-105$&$63$&$-84$&$-14$&$56$&$-27$&$-14$&$20$&$-3$&$2$&$-4$&$1$&\\ \hline
$4(7,6)$&$14$&$21$&$-28$&$70$&$-28$&$84$&$14$&$-56$&$45$&$7$&$-31$&$11$&$-2$&$9$&$-6$&$1$\\ \hline
\end{tabular}
\bigskip
\caption{The polynomials $c^+_{(\la,\mu)}$ of the congruence class $\#=1$ of degrees $\leq7$. In the first column the labels $(\la,\mu)$ are shown. The constants 2 and 4 should be understood as multiplying the functions $c^+_{(\la,\mu)}$, making all the entries of the table integer.}
\label{cpl}
\end{center}
\end{table}

\begin{table}[!h]
\begin{center}
\small
\begin{tabular}{|c|@{\,}r@{\,}|@{\,}r@{\,}|@{\,}r@{\,}|@{\,}r@{\,}|@{\,}r@{\,}|@{\,}r@{\,}|@{\,}r@{\,}|@{\,}r@{\,}|@{\,}r@{\,}|@{\,}r@{\,}|@{\,}r@{\,}|@{\,}r@{\,}|}
\hline
$\frac{1}{U}c^-$&\footnotesize{$X$}&\footnotesize{$XY$}&\footnotesize{$X^3$}&\footnotesize{$XY^2$}&\footnotesize{$X^3Y$}&\footnotesize{$XY^3$}&\footnotesize{$X^5$}&\footnotesize{$X^3Y^2$}&\footnotesize{$XY^4$}&\footnotesize{$X^5Y$}&\footnotesize{$X^3Y^3$}&\footnotesize{$XY^5$}\\ \hline
$2(2,0)$&$1$&&&&&&&&&&&\\ \hline
$4(3,1)$&$0$&$1$&&&&&&&&&&\\ \hline
$2(4,0)$&$-4$&$-2$&$1$&&&&&&&&&\\ \hline
$4(4,2)$&$6$&$4$&$-2$&$1$&&&&&&&&\\ \hline
$4(5,1)$&$0$&$-5$&$0$&$-2$&$1$&&&&&&&\\ \hline
$4(5,3)$&$0$&$10$&$0$&$6$&$-3$&$1$&&&&&&\\ \hline
$2(6,0)$&$9$&$12$&$-6$&$3$&$-4$&$0$&$1$&&&&&\\ \hline
$4(6,2)$&$-16$&$-24$&$12$&$-12$&$8$&$-2$&$-2$&$1$&&&&\\ \hline
$4(6,4)$&$10$&$20$&$-10$&$21$&$-8$&$8$&$2$&$-4$&$1$&&&\\ \hline
$4(7,1)$&$0$&$14$&$0$&$14$&$-7$&$3$&$0$&$-4$&$0$&$1$&&\\ \hline
$4(7,3)$&$0$&$-35$&$0$&$-42$&$21$&$-16$&$0$&$12$&$-2$&$-3$&$1$&\\ \hline
$4(7,5)$&$0$&$35$&$0$&$56$&$-28$&$36$&$0$&$-20$&$10$&$5$&$-5$&$1$\\ \hline
\end{tabular}
\bigskip
\caption{The polynomials $U^{-1}c^-_{(\la,\mu)}$ of the congruence class $\#=0$ of degrees $\leq6$. In the first column the labels $(\la,\mu)$ are shown. The constants 2 and 4 should be understood as multiplying the functions $U^{-1}c^-_{(\la,\mu)}$, making all the entries of the table integer.}
\label{cms}
\end{center}
\end{table}

\begin{table}[!h]
\begin{center}
\small 
\begin{tabular}{|c|@{\,}r@{\,}|@{\,}r@{\,}|@{\,}r@{\,}|@{\,}r@{\,}|@{\,}r@{\,}|@{\,}r@{\,}|@{\,}r@{\,}|@{\,}r@{\,}|@{\,}r@{\,}|@{\,}r@{\,}|@{\,}r@{\,}|@{\,}r@{\,}|@{\,}r@{\,}|@{\,}r@{\,}|@{\,}r@{\,}|@{\,}r@{\,}|}
\hline
$\frac{1}{U}c^-$&\footnotesize{$1$}&\footnotesize{$Y$}&\footnotesize{$X^2$}&\footnotesize{$Y^2$}&\footnotesize{$X^2Y$}&\footnotesize{$Y^3$}&\footnotesize{$X^4$}&\footnotesize{$X^2Y^2$}&\footnotesize{$Y^4$}&\footnotesize{$X^4Y$}&\footnotesize{$X^2Y^3$}&\footnotesize{$Y^5$}&\footnotesize{$X^6$}&\footnotesize{$X^4Y^2$}&\footnotesize{$X^2Y^4$}&\footnotesize{$Y^6$}\\ \hline
$2(1,0)$&$1$&&&&&&&&&&&&&&&\\ \hline
$4(2,1)$&$2$&$1$&&&&&&&&&&&&&&\\ \hline
$2(3,0)$&$-3$&$-1$&$1$&&&&&&&&&&&&&\\ \hline
$4(3,2)$&$6$&$5$&$-2$&$1$&&&&&&&&&&&&\\ \hline
$4(4,1)$&$-2$&$-4$&$0$&$-1$&$1$&&&&&&&&&&&\\ \hline
$4(4,3)$&$6$&$14$&$-2$&$7$&$-3$&$1$&&&&&&&&&&\\ \hline
$2(5,0)$&$5$&$5$&$-5$&$1$&$-3$&$0$&$1$&&&&&&&&&\\ \hline
$4(5,2)$&$-10$&$-15$&$10$&$-7$&$6$&$-1$&$-2$&$1$&&&&&&&&\\ \hline
$4(5,4)$&$10$&$30$&$-10$&$27$&$-11$&$9$&$2$&$-4$&$1$&&&&&&&\\ \hline
$4(6,1)$&$2$&$9$&$0$&$6$&$-6$&$1$&$0$&$-3$&$0$&$1$&&&&&&\\ \hline
$4(6,3)$&$-6$&$-29$&$2$&$-27$&$18$&$-9$&$0$&$9$&$-1$&$-3$&$1$&&&&&\\ \hline
$4(6,5)$&$10$&$55$&$-10$&$77$&$-36$&$44$&$2$&$-24$&$11$&$5$&$-5$&$1$&&&&\\ \hline
$2(7,0)$&$-7$&$-14$&$14$&$-7$&$21$&$-1$&$-7$&$6$&$0$&$-5$&$0$&$0$&$1$&&&\\ \hline
$4(7,2)$&$14$&$35$&$-28$&$28$&$-42$&$9$&$14$&$-19$&$1$&$10$&$-3$&$0$&$-2$&$1$&&\\ \hline
$4(7,4)$&$-14$&$-56$&$28$&$-77$&$49$&$-44$&$-14$&$40$&$-11$&$-10$&$12$&$-1$&$2$&$-4$&$1$&\\ \hline
$4(7,6)$&$14$&$91$&$-28$&$182$&$-84$&$156$&$14$&$-96$&$65$&$17$&$-41$&$13$&$-2$&$9$&$-6$&$1$\\ \hline
\end{tabular}
\bigskip
\caption{Polynomials $U^{-1}c^-_{(\la,\mu)}$ of the congruence class $\#=1$ of degrees $\leq6$. Labels $(\la,\mu)$ are given in the first column. Constants 2 and 4 should be understood as multiplying functions $U^{-1}c^-_{(\la,\mu)}$, making all entries of the table integer.}
\label{cml}
\end{center}
\end{table}

\begin{table}[!h]
\begin{center}
\small 
\begin{tabular}{|c|@{\,}r@{\,}|@{\,}r@{\,}|@{\,}r@{\,}|@{\,}r@{\,}|@{\,}r@{\,}|@{\,}r@{\,}|@{\,}r@{\,}|@{\,}r@{\,}|@{\,}r@{\,}|@{\,}r@{\,}|@{\,}r@{\,}|@{\,}r@{\,}|@{\,}r@{\,}|@{\,}r@{\,}|@{\,}r@{\,}|@{\,}r@{\,}|}
\hline
$\frac{1}{V}s^+$&\footnotesize{$1$}&\footnotesize{$Y$}&\footnotesize{$X^2$}&\footnotesize{$Y^2$}&\footnotesize{$X^2Y$}&\footnotesize{$Y^3$}&\footnotesize{$X^4$}&\footnotesize{$X^2Y^2$}&\footnotesize{$Y^4$}&\footnotesize{$X^4Y$}&\footnotesize{$X^2Y^3$}&\footnotesize{$Y^5$}&\footnotesize{$X^6$}&\footnotesize{$X^4Y^2$}&\footnotesize{$X^2Y^4$}&\footnotesize{$Y^6$}\\ \hline
$2(1,1)$&$1$&&&&&&&&&&&&&&&\\ \hline
$2(2,2)$&$0$&$1$&&&&&&&&&&&&&&\\ \hline
$4(3,1)$&$-2$&$-2$&$1$&&&&&&&&&&&&&\\ \hline
$2(3,3)$&$1$&$2$&$-1$&$1$&&&&&&&&&&&&\\ \hline
$4(4,2)$&$0$&$-4$&$0$&$-2$&$1$&&&&&&&&&&&\\ \hline
$2(4,4)$&$0$&$4$&$0$&$4$&$-2$&$1$&&&&&&&&&&\\ \hline
$4(5,1)$&$2$&$6$&$-3$&$2$&$-4$&$0$&$1$&&&&&&&&&\\ \hline
$4(5,3)$&$-2$&$-8$&$4$&$-8$&$4$&$-2$&$-1$&$1$&&&&&&&&\\ \hline
$2(5,5)$&$1$&$6$&$-3$&$11$&$-4$&$6$&$1$&$-3$&$1$&&&&&&&\\ \hline
$4(6,2)$&$0$&$6$&$0$&$8$&$-4$&$2$&$0$&$-4$&$0$&$1$&&&&&&\\ \hline
$4(6,4)$&$0$&$-12$&$0$&$-22$&$11$&$-12$&$0$&$8$&$-2$&$-2$&$1$&&&&&\\ \hline
$2(6,6)$&$0$&$9$&$0$&$24$&$-12$&$22$&$0$&$-12$&$8$&$3$&$-4$&$1$&&&&\\ \hline
$4(7,1)$&$-2$&$-12$&$6$&$-10$&$20$&$-2$&$-5$&$9$&$0$&$-6$&$0$&$0$&$1$&&&\\ \hline
$4(7,3)$&$2$&$14$&$-7$&$22$&$-20$&$12$&$5$&$-14$&$2$&$6$&$-4$&$0$&$-1$&$1$&&\\ \hline
$4(7,5)$&$-2$&$-18$&$9$&$-48$&$24$&$-44$&$-6$&$30$&$-16$&$-6$&$12$&$-2$&$1$&$-3$&$1$&\\ \hline
$2(7,7)$&$1$&$12$&$-6$&$46$&$-20$&$62$&$5$&$-39$&$37$&$6$&$-24$&$10$&$-1$&$6$&$-5$&$1$\\ \hline
\end{tabular}
\bigskip
\caption{Ppolynomials $V^{-1}s^+_{(\la,\mu)}$ of the congruence class $\#=0$ of degrees $\leq6$. Labels $(\la,\mu)$ are given in the first column. Constants 2 and 4 should be understood as multiplying the functions $V^{-1}s^+_{(\la,\mu)}$, making all entries of the table integer.}
\label{sps}
\end{center}
\end{table}

\begin{table}[!h]
\begin{center}
\small
\begin{tabular}{|c|@{\,}r@{\,}|@{\,}r@{\,}|@{\,}r@{\,}|@{\,}r@{\,}|@{\,}r@{\,}|@{\,}r@{\,}|@{\,}r@{\,}|@{\,}r@{\,}|@{\,}r@{\,}|@{\,}r@{\,}|@{\,}r@{\,}|@{\,}r@{\,}|}
\hline
$\frac{1}{V}s^+$&\footnotesize{$X$}&\footnotesize{$XY$}&\footnotesize{$X^3$}&\footnotesize{$XY^2$}&\footnotesize{$X^3Y$}&\footnotesize{$XY^3$}&\footnotesize{$X^5$}&\footnotesize{$X^3Y^2$}&\footnotesize{$XY^4$}&\footnotesize{$X^5Y$}&\footnotesize{$X^3Y^3$}&\footnotesize{$XY^5$}\\ \hline
$4(2,1)$&$1$&&&&&&&&&&&\\ \hline
$4(3,2)$&$-1$&$1$&&&&&&&&&&\\ \hline
$4(4,1)$&$-2$&$-3$&$1$&&&&&&&&&\\ \hline
$4(4,3)$&$2$&$1$&$-1$&$1$&&&&&&&&\\ \hline
$4(5,2)$&$1$&$-3$&$0$&$-3$&$1$&&&&&&&\\ \hline
$4(5,4)$&$-2$&$3$&$1$&$3$&$-2$&$1$&&&&&&\\ \hline
$4(6,1)$&$3$&$12$&$-4$&$5$&$-5$&$0$&$1$&&&&&\\ \hline
$4(6,3)$&$-3$&$-9$&$4$&$-9$&$5$&$-3$&$-1$&$1$&&&&\\ \hline
$4(6,5)$&$3$&$3$&$-4$&$8$&$-2$&$5$&$1$&$-3$&$1$&&&\\ \hline
$4(7,2)$&$-1$&$6$&$0$&$15$&$-5$&$5$&$0$&$-5$&$0$&$1$&&\\ \hline
$4(7,4)$&$2$&$-9$&$-1$&$-24$&$10$&$-15$&$0$&$10$&$-3$&$-2$&$1$&\\ \hline
$4(7,6)$&$-3$&$6$&$4$&$16$&$-10$&$17$&$-1$&$-9$&$7$&$3$&$-4$&$1$\\ \hline
\end{tabular}
\bigskip
\caption{Polynomials $V^{-1}s^+_{(\la,\mu)}$ of the congruence class $\#=1$ of degrees $\leq6$. Labels $(\la,\mu)$ are given in the first column. The constant 4 should be understood as multiplying the functions $V^{-1}s^+_{(\la,\mu)}$, making all entries of the table integer.}
\label{spl}
\end{center}
\end{table}

\begin{table}[!h]
\begin{center}
\small
\begin{tabular}{|c|@{\,}r@{\,}|@{\,}r@{\,}|@{\,}r@{\,}|@{\,}r@{\,}|@{\,}r@{\,}|@{\,}r@{\,}|@{\,}r@{\,}|@{\,}r@{\,}|@{\,}r@{\,}|}
\hline
$\frac{1}{W}s^-$&\footnotesize{$X$}&\footnotesize{$XY$}&\footnotesize{$X^3$}&\footnotesize{$XY^2$}&\footnotesize{$X^3Y$}&\footnotesize{$XY^3$}&\footnotesize{$X^5$}&\footnotesize{$X^3Y^2$}&\footnotesize{$XY^4$}\\ \hline
$4(3,1)$&$1$&&&&&&&&\\ \hline
$4(4,2)$&$0$&$1$&&&&&&&\\ \hline
$4(5,1)$&$-3$&$-2$&$1$&&&&&&\\ \hline
$4(5,3)$&$2$&$2$&$-1$&$1$&&&&&\\ \hline
$4(6,2)$&$0$&$-4$&$0$&$-2$&$1$&&&&\\ \hline
$4(6,4)$&$0$&$5$&$0$&$4$&$-2$&$1$&&&\\ \hline
$4(7,1)$&$6$&$10$&$-5$&$3$&$-4$&$0$&$1$&&\\ \hline
$4(7,3)$&$-5$&$-10$&$5$&$-8$&$4$&$-2$&$-1$&$1$&\\ \hline
$4(7,5)$&$3$&$8$&$-4$&$12$&$-4$&$6$&$1$&$-3$&$1$\\ \hline
\end{tabular}
\bigskip
\caption{Polynomials $W^{-1}s^-_{(\la,\mu)}$ of the congruence class $\#=0$ of degrees $\leq5$. Labels $(\la,\mu)$ are given in the first column. The constant 4 should be understood as multiplying the functions $W^{-1}s^-_{(\la,\mu)}$, making all the entries of the table integer.}
\label{sms}
\end{center}
\end{table}

\begin{table}[!h]
\begin{center}
\small 
\begin{tabular}{|c|@{\,}r@{\,}|@{\,}r@{\,}|@{\,}r@{\,}|@{\,}r@{\,}|@{\,}r@{\,}|@{\,}r@{\,}|@{\,}r@{\,}|@{\,}r@{\,}|@{\,}r@{\,}|@{\,}r@{\,}|@{\,}r@{\,}|@{\,}r@{\,}|}
\hline
$\frac{1}{W}s^-$&\footnotesize{$1$}&\footnotesize{$Y$}&\footnotesize{$X^2$}&\footnotesize{$Y^2$}&\footnotesize{$X^2Y$}&\footnotesize{$Y^3$}&\footnotesize{$X^4$}&\footnotesize{$X^2Y^2$}&\footnotesize{$Y^4$}&\footnotesize{$X^4Y$}&\footnotesize{$X^2Y^3$}&\footnotesize{$Y^5$}\\ \hline
$4(2,1)$&$1$&&&&&&&&&&&\\ \hline
$4(3,2)$&$1$&$1$&&&&&&&&&&\\ \hline
$4(4,1)$&$-2$&$-1$&$1$&&&&&&&&&\\ \hline
$4(4,3)$&$2$&$3$&$-1$&$1$&&&&&&&&\\ \hline
$4(5,2)$&$-1$&$-3$&$0$&$-1$&$1$&&&&&&&\\ \hline
$4(5,4)$&$2$&$7$&$-1$&$5$&$-2$&$1$&&&&&&\\ \hline
$4(6,1)$&$3$&$4$&$-4$&$1$&$-3$&$0$&$1$&&&&&\\ \hline
$4(6,3)$&$-3$&$-7$&$4$&$-5$&$3$&$-1$&$-1$&$1$&&&&\\ \hline
$4(6,5)$&$3$&$13$&$-4$&$16$&$-6$&$7$&$1$&$-3$&$1$&&&\\ \hline
$4(7,2)$&$1$&$6$&$0$&$5$&$-5$&$1$&$0$&$-3$&$0$&$1$&&\\ \hline
$4(7,4)$&$-2$&$-13$&$1$&$-16$&$10$&$-7$&$0$&$6$&$-1$&$-2$&$1$&\\ \hline
$4(7,6)$&$3$&$22$&$-4$&$40$&$-18$&$29$&$1$&$-15$&$9$&$3$&$-4$&$1$\\ \hline
\end{tabular}
\bigskip
\caption{Polynomials $W^{-1}s^-_{(\la,\mu)}$ of the congruence class $\#=1$ of degrees $\leq5$. Labels $(\la,\mu)$ are given in the first column. The constant 4 should be understood as multiplying the functions $W^{-1}s^-_{(\la,\mu)}$, making all the entries of the table integer.}
\label{sml}
\end{center}
\end{table}

\end{document}